\documentclass[12pt]{spieman}
\usepackage{color}
\usepackage{amsmath,amsfonts,amssymb}
\usepackage{graphicx}
\usepackage{setspace}
\usepackage{tocloft}

\newcommand{\mycitet}[1]{Ref.~\citenum{#1}}
\newcommand{\mycitep}[1]{\cite{#1}}
\newcommand{\dd}[2]{\frac{\partial #1}{\partial #2}}

\title{High-order wavefront sensing and control for the Roman Coronagraph Instrument (CGI): architecture and measured performance}
\author[a]{Eric Cady}
\author[a]{Nicholas Bowman}
\author[b]{Alexandra Z. Greenbaum}
\author[b]{James G. Ingalls}
\author[a]{Brian Kern}
\author[a]{John Krist}
\author[a]{David Marx}
\author[a]{Ilya Poberezhskiy}
\author[a]{A J Eldorado Riggs}
\author[a]{Garreth Ruane}
\author[a]{Byoung-Joon Seo}
\author[a]{Fang Shi}
\author[a]{Hanying Zhou}
\affil[a]{Jet Propulsion Laboratory, California Institute of Technology, 4800 Oak Grove Drive, Pasadena, CA, USA 91109}
\affil[b]{IPAC, California Institute of Technology, 1200 E California Blvd, Pasadena, CA, USA 91125}

\cftpagenumbersoff{figure}
\cftpagenumbersoff{table} 
\begin{document} 
\maketitle

\let\thefootnote\relax\footnote{\copyright 2024 California Institute of Technology. Government sponsorship acknowledged.}

\begin{abstract}
The Nancy Grace Roman Space Telescope (``Roman'') is a 2.4m space telescope scheduled for a 2026 launch.  The Coronagraph Instrument (CGI) on Roman is a technology-demonstration instrument with a coronagraph and, for the first time in space, deformable mirrors and active wavefront control.  This paper walks through the algorithmic and system-level architecture of the HOWFSC implementation for CGI, including the use of ground-in-the-loop (GITL) operations to support computationally-expensive operations, and reports on instrument performance measured during thermal vacuum testing in instrument integration and test.  CGI achieved better than $5\times10^{-8}$ total raw contrast with two independent coronagraph architectures covering 3-9 and 6-20 $\lambda/D$ between them and a $360^{\circ}$ dark hole on each.  The contrast limits appear to be driven by time available for testing, and do not appear to represent a floor in the achievable performance of CGI in flight.
\end{abstract}

\keywords{Roman Space Telescope, CGI, coronagraphy, wavefront sensing, wavefront control, HOWFSC}

{\noindent \footnotesize\textbf{*} Eric Cady,  \linkable{eric.j.cady@jpl.nasa.gov}}

\section{Introduction} \label{sec:intro}

\noindent The Nancy Grace Roman Space Telescope (``Roman'' throughout this work) is a 2.4m flagship-class space telescope scheduled for launch in October 2026.  Roman has a primary Wide Field Instrument (WFI) for wide-field infrared survey data collection in support of cosmology and detecting microlensing exoplanets, and it has a secondary Coronagraph Instrument (CGI\mycitep{Men22, Pob22}) which is a class-D technology demonstration instrument designed to demonstrate technologies applicable to future space missions\mycitep{Fei24} capable of imaging Earth-size exoplanets with coronagraphy.  (See \mycitet{Men22, Pob22} for more detailed overviews of CGI; see \mycitet{Gal23} for more information on coronagraphy in general.)

On the hardware side, these technological elements include coronagraph masks (both focal- and pupil-plane), low-order wavefront sensors, 2 deformable mirrors (DMs) with $48\times48$ grids of actuators apiece, and detectors with photon-counting capabilities.  But from a software and operational perspective, CGI is also a pathfinder for high-order wavefront sensing and control (HOWFSC), which uses data collected with the CGI science camera (EXCAM) to drive the DMs, in order to control starlight in focal-plane regions of interest (``dark holes'').

This fine control is necessary as CGI has a single Level 1 requirement, Threshold Technology Requirement 5 (TTR5): \textit{Roman shall be able to measure (using CGI), with SNR $\geq 5$, the brightness of an astrophysical point source located between 6 and 9 $\lambda/D$ from an adjacent star with a $V_{AB}$ magnitude $\leq 5$, with a flux ratio $\geq 1\times 10^{-7}$; the bandpass shall have a central wavelength $\leq 600$ nm and a bandwidth $\geq 10$\%}.  Subsequent decomposition of this requirement through an error budget and a concept of operations (see \mycitet{Nem23} for details on the error budgeting) led to the following constraints on contrast:
\begin{itemize}
\item Mean static coherent raw contrast over the 6-9 $\lambda/D$ annulus with CGI's Band 1: $\leq 5\times10^{-8}$
\item Mean static incoherent raw contrast over the 6-9 $\lambda/D$ annulus with CGI's Band 1: $\leq 1\times10^{-7}$
\end{itemize}
The definition for contrast here is as follows, reproduced from \mycitet{Kri23}: ``\textit{contrast at a given location is the ratio of the per-pixel field intensity divided by the peak pixel intensity of the star if it was centered at that position}.''  Coherent contrast is the portion of the contrast from speckles which are coherent with the light from the target star and may be modulated with the DMs; incoherent contrast is the portion of the contrast from sources that are not coherent with the light from the target, such as other astrophysical sources or stray light.

These values are orders of magnitude better than the state of the art capability in ground- or space-based coronagraphy \mycitep{Cou23, Gir22}, and cannot be achieved with high-quality optics and alignment on their own.  To reach these levels, we must use active sensing and control to measure the residuals and compensate for them across a dark hole.  DMs are an enabling technology for this, and while DMs have been tested in high-altitude balloons (PICTURE-C\mycitep{Men23}) and cubesats (DeMI\mycitep{Mor22}), CGI will be the first space telescope to use them for wavefront control in combination with a coronagraph.

The CGI HOWFSC system at this time has been designed, implemented, and tested in relevant vacuum environment on the instrument.  This paper will cover:
\begin{itemize}
\item The algorithmic architecture of CGI HOWFSC (algorithm selection and implementation)
\item The system-level architecture of CGI HOWFSC (splitting computation between flight and ground software)
\item Performance results from testing HOWFSC on the flight instrument during the instrument integration and test (II\&T) thermal vacuum (TVAC) campaign
\end{itemize}

\section{CGI High-Order Wavefront Sensing}

\noindent The following subsections will walk through the HOWFSC approach used by CGI.  CGI algorithms have used broadly the same wavefront estimation and control as was used during CGI Technology Milestone testing during the 2014-2017 period \mycitep{Cad16, Seo16, Cad17, Seo17, Mar17} in the High Contrast Imaging Testbed (HCIT) at JPL.  This was an intentional choice, as the purpose of the Milestone testing was to demonstrate that the algorithms would allow us to achieve contrasts consistent with CGI requirements given the limitations of the Roman pupil, which had a larger secondary diameter and thicker spiders than a pupil selected with coronagraphy in mind.  We chose, wherever possible, to bring forward existing algorithmic approaches rather than invent new ones.  This both saved implementation time and addressed concerns about algorithm risk.  CGI HOWFSC did update on an as-needed basis in order to meet requirements: for example, incorporating the fast Jacobian calculation approach from the FALCO \mycitep{Rig18} wavefront control software package helped enable per-iteration Jacobian calculation (see Sec. \ref{subsec:jac}) with no downtime.


The algorithmic approach has two parts: a wavefront estimation based around pairwise probing\mycitep{Giv11}, where pairs of DM ``probes'' create structured modulations to allow amplitude and phase to be extracted from intensity, and a wavefront control step based around electric field conjugation \mycitep{Giv07}, which uses a model of the optical system to set up and solve a least-squares problem for the measured electric field.

\subsection{Probe Shapes} \label{subsec:probes}

\noindent The default CGI HOWFSC probe is an extension on the sinc-sinc-sin approach suggested in \mycitet{Giv07}; the pair of sincs in the pupil plane is used to create a rectangular region of modulation in the focal plane, and the sine creates two copies of the rectangle on either side of the PSF, with a phase difference between them.  A list of inputs to the probe shape definition is given in Table \ref{tab:probe}.

\begin{table}[ht]
\caption{Inputs used to define a CGI HOWFSC probe pattern.} 
\label{tab:probe}
\begin{center}       
\begin{tabular}{|p{0.1\linewidth}|p{0.7\linewidth}|}
\hline
Variable & Definition \\
\hline
$N_{act}$ & Number of actuators along one side of the DM.  (Assumes square DM.) \\
\hline
$D_{act}$ & Diameter of pupil, in number of actuators.  May be non-integer.  Used to get $\lambda/D$ right.\\
\hline
$x_{c}$ & Number of actuators to offset the center of the DM pattern along the positive x-axis, as seen from the camera.  Negative and fractional inputs are acceptable. \\
\hline
$y_{c}$ & Number of actuators to offset the center of the DM pattern along the positive y-axis, as seen from the camera.  Negative and fractional inputs are acceptable. \\
\hline
$\theta$ & Clockwise angle to rotate the DM pattern about its center. \\
\hline
$h$ & Height of sinc peak; actual shape may not reach this value depending on centration \\
\hline
$\xi_{min}$ & Minimum $\lambda/D$ along the x-axis in the focal plane for the probe rectangle.  Must be less than $\xi_{max}$. \\
\hline
$\xi_{max}$ & Maximum $\lambda/D$ along the x-axis in the focal plane for the probe rectangle. \\
\hline
$\eta_{min}$ & Minimum $\lambda/D$ along the y-axis in the focal plane for the probe rectangle.  Must be less than $\eta_{max}$. \\
\hline
$\eta_{max}$ & Maximum $\lambda/D$ along the y-axis in the focal plane for the probe rectangle. \\
\hline
$\phi$ & Phase angle to shift this particular probe; at phase = $0$ the modulation will be a sine, and at phase = $\pi/2$ it will be a cosine. \\
\hline
\end{tabular}
\end{center}
\end{table} 
In practice, creating a probe may require iteration:
\begin{itemize}
\item $x_{c}$ and $y_{c}$ should be checked by overlaying the DM grid on the OTA pupil, shaped pupil, and Lyot stop to be used with the probe pair, and verify that the core is unobstructed by pupil-plane artifacts.  $x_{c}$ and $y_{c}$ should be chosen to keep the core unblocked; they do not affect the region of modulation.
\item $h$ will need to be tuned so that the probe intensity is at a known level.  CGI uses scaling factors when applying the probe so the same probe pattern can create stronger or weaker modulations, but the intensity associated with the unscaled probe must be known to make use of this.  In CGI ground software, we use the HOWFSC optical model (Sec. \ref{subsec:model}) to propagate probed electric-fields, and use the model-based amplitude estimate (Sec. \ref{subsec:pp}) to determine probe intensity in a region of interest and set $h$ appropriately.
\end{itemize}

\begin{figure}
\centering
\includegraphics[width=1\textwidth]{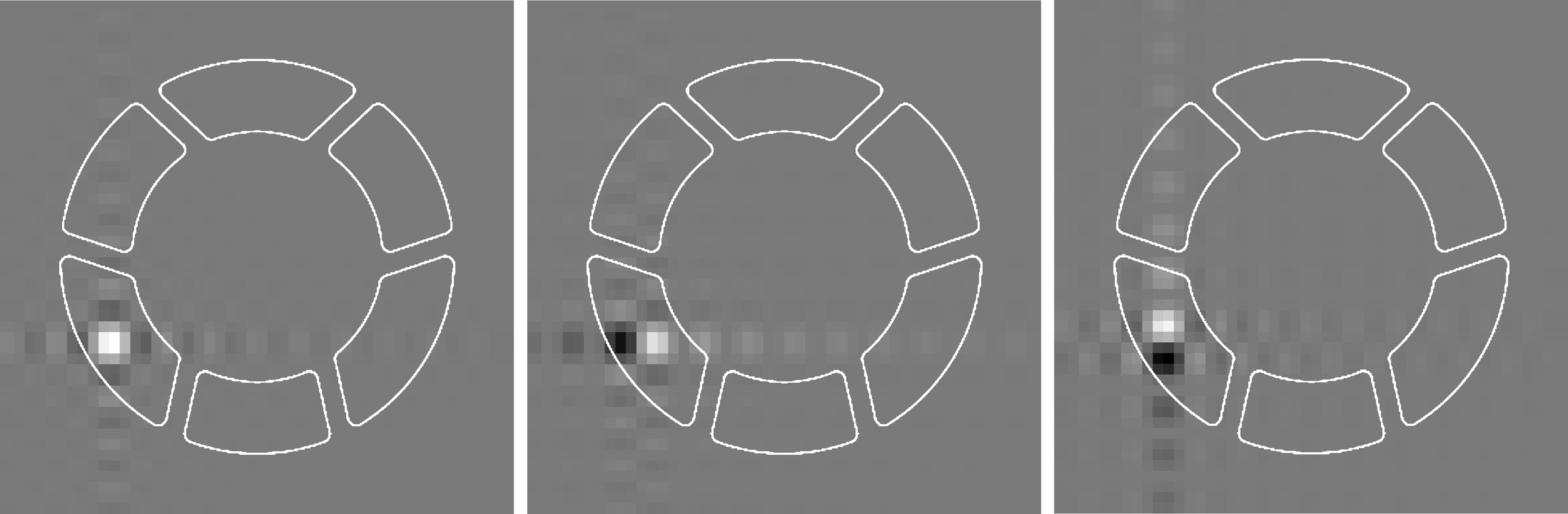}
\caption{Three NFOV HLC sinc-sinc-sin probes used during CGI TVAC operations.  The limiting pupil-plane aperture (HLC Lyot stop) is overlaid.}
\label{fig:3probe}
\end{figure}

CGI HOWFSC uses three probe pairs.  Our baseline is to use one cosine ($\phi = \pi/2, \theta = 0$) and two sines with rectangles oriented left-right or up-down ($\phi = 0, \theta = 0$ and $\pi/2$).  For a $360^{\circ}$ dark hole, the CGI baseline for imaging, we select $\xi_{max} = \eta_{max} = -\eta_{min}$ to create a square region of modulation, and select $\xi_{min} = 0$, so there is no gap between the rectangles.  We select $\xi_{max}$ to slightly oversize the dark hole region, so that if any modulation is possible in the wings leaking in around the outer working angle, we exercise it.  An example of three probe pairs, used during TVAC with the narrow field-of-view (NFOV) hybrid Lyot coronagraph (HLC) configuration, is shown in Fig. \ref{fig:3probe}.

The probe definition $P$ itself, in units of surface height, can be written as:
\begin{align}
\mathrm{Define\,a\,2D\,grid\,on\,} x_{0} \mathrm{\,and\,} y_{0} \mathrm{\,with:} \nonumber & \\
x_{0} &= [\frac{-N_{act}}{2} + \frac{1}{2}, \hdots, \frac{N_{act}}{2} - \frac{1}{2}] \\
y_{0} &= [\frac{-N_{act}}{2} + \frac{1}{2}, \hdots, \frac{N_{act}}{2} - \frac{1}{2}] \\
\mathrm{For\,each\,grid\,point,\,compute:}
\begin{bmatrix}
x  \\ y
\end{bmatrix}
&= 
\begin{bmatrix}
\cos{\theta} & -\sin{\theta} \\ \sin{\theta} & \cos{\theta}
\end{bmatrix}
\begin{bmatrix}
x_{0} - x_{c}  \\ y_{0} - y_{c}
\end{bmatrix} \\
W_{x} &= \frac{D_{act}}{(\xi_{max} - \xi_{min})} \\
W_{y} &= \frac{D_{act}}{(\eta_{max} - \eta_{min})} \\
f_{x} &= \frac{(\xi_{max} + \xi_{min})}{2} \\
f_{y} &= \frac{(\eta_{max} + \eta_{min})}{2}
\end{align}
\begin{align}
P &= \frac{2 h}{W_{x} W_{y}} \frac{\sin{\pi x/W_{x}}}{\pi x/W_{x}} \frac{\sin{\pi y/W_{y}}}{\pi y/W_{y}} \sin{\left(2 \pi (x f_{x} + y f_{y})/D_{act} + \phi \right)}
\end{align}

We note for completeness that CGI probes are implemented as ``relative DM settings'', $48 \times 48$ arrays which can be added to an existing DM setting after optionally being multiplied by a scale factor.  By default, these probe settings represent probe normalized intensities of $10^{-5}$ of the PSF peak, averaged across the entire dark hole region, when used with a scale factor of $\pm1$.  There are no practical limitations to using a different probe for HOWFSC beyond the effort to create the new $48 \times 48$ array, and to update a ground-based configuration file to document the correct intensity if that probe intensity was not scaled to $10^{-5}$.  No part of HOWFSC operations relies on the probes having any particular shape, including sinc-sinc-sin; HOWFSC will reconstruct what was actually applied from the DM settings being telemetered.  

The choice of sinc-sinc-sin for the CGI baseline does have some theoretical basis in noise optimality (see analysis in \mycitet{Gro15}), but inertia also played a role: the probes had been historically effective in HCIT and used throughout CGI Milestone testing\mycitep{Cad16, Seo16, Cad17, Seo17}.  We note that other high-contrast facilities have more recently used or tested other probe patterns, such as single-actuator probes\mycitep{Ahn23}, double-actuator probes\mycitep{Pot20}, power-law probes\mycitep{Van24}, and probes optimized using the system model through the Jacobian\mycitep{Red21}.  Now that the CGI software functionality has matured, there likely exists unexplored parameter space on CGI to improve our wavefront estimation by updating the relative DM settings we use for probes.  As these can be updated without a new software release, it does suggest this area of investigation would have a relatively high return on investment.

\subsection{Probe Phase and Amplitude Estimation} \label{subsec:pp}

For the purposes of wavefront estimation, the intensity pattern from the observation of an occulted star can be decomposed into the intensity pattern from the diffracted starlight itself ($I_{coh}$, which is related to the electric field from the star $E_{0}$) and the sum of any incoherent backgrounds (astrophysical, stray light, etc.) $I_{inc}$ which do not interfere with the starlight and cannot be affected by DM probing.  Extracting $E_0$ from intensity measurement requires knowledge of the probe signal used to produce the modulation; this section describes how CGI estimates the probe phase and amplitude.

\noindent Using the convention in \mycitet{Giv11}, pairwise probing creates an approximate probe field $\Delta p$ with a relationship to intensity images as:
\begin{align}
I_{0} &= I_{coh}  + I_{inc} = \left|C[E_{0}]\right|^2  + I_{inc} \label{eq:iinc} \\
I_{+} &= \left|C[E_{0} e^{i \Delta \psi}]\right|^2  + I_{inc} \approx \left|C[E_{0}]  + i C[E_{0} \Delta \psi]\right|^2 + I_{inc} \equiv  \left|C[E_{0}]  + i \Delta p\right|^2 + I_{inc}\\
I_{-} &= \left|C[E_{0} e^{-i \Delta \psi}]\right|^2 + I_{inc} \approx \left|C[E_{0}]  - i C[E_{0} \Delta \psi]\right|^2 + I_{inc}\equiv  \left|C[E_{0}]  - i \Delta p\right|^2 + I_{inc}
\end{align}
for some coronagraph propagation operator $C$ which starts at the DM and propagates to the final focal plane.

Implicitly, we have defined
\begin{align}
\mathrm{Solving\,from}\, I_{+}: \Delta p &= \frac{C[E_{0} e^{i \Delta \psi}] - C[E_{0}]}{i} \label{eq:dp1}\\
\mathrm{Solving\,from}\, I_{-}: \Delta p &= \frac{C[E_{0}] - C[E_{0} e^{-i \Delta \psi}]}{i} \label{eq:dp2}
\end{align}
on the assumption that the two are the same.  We want to estimate probe phase (angle of $\Delta p$) while reconciling these two.  If we have a model for $C$ (Sec. \ref{subsec:model}), easiest way is to average the two:
\begin{align}
\Delta p_{\mathrm{effective}} &= \frac{\left(C[E_{0} e^{i \Delta \psi}] - C[E_{0}]\right) + \left(C[E_{0}] - C[E_{0} e^{-i \Delta \psi}]\right)}{2 i} \\
&= \frac{\left(C[E_{0} e^{i \Delta \psi}] - C[E_{0} e^{-i \Delta \psi}]\right)}{2 i}
\end{align}
This will be the same result as Eqs. \ref{eq:dp1} and \ref{eq:dp2} if the positive and negative probe effect on the field is in fact the same, and will split the difference otherwise.  To get the model-based probe phase and probe amplitude, take the argument and absolute value of $\Delta p_{\mathrm{effective}}$.  $\Delta p_{\mathrm{effective}}$ is complex-valued.  

CGI HOWFSC extracts probe phase in this manner, but uses empirical estimation for amplitude rather than the model-based estimate for HOWFSC.  (Empirical phase would be preferred as well, but it cannot be derived directly in pairwise probing, hence the above model-based calculation.)
\begin{align}
I_{p} &= \frac{I_{+} + I_{-}}{2} - I_{0} = \frac{\left|C[E_{0} e^{i \Delta \psi}]\right|^2  + \left|C[E_{0} e^{-i \Delta \psi}]\right|^2}{2} - \left|C[E_{0}]\right|^2 \label{eq:ip} \\
&\approx \frac{\left|C[E_{0}]  + i \Delta p\right|^2 + \left|C[E_{0}]  - i \Delta p\right|^2}{2} - \left|C[E_{0}]\right|^2 \\
&\equiv \frac{2\left|C[E_{0}]\right|^2 + 2\left|\Delta p\right|^2}{2} - \left|C[E_{0}]\right|^2 = \left|\Delta p\right|^2 \\
\Rightarrow |\Delta p| &\approx \sqrt{\frac{I_{+} + I_{-}}{2} - I_{0}} \label{eq:dpa}
\end{align}
Model-based amplitudes are still used for adjusting the amplitude of HOWFSC probes ($h$) to give an average normalized intensity across the dark hole of $10^{-5}$ when preparing the associated DM settings.  This is done running the model against itself without requiring instrument data to be collected.

\subsection{Estimation Calculations} \label{subsec:est}

\noindent CGI HOWFSC uses the estimation approach outlined in \mycitet{Giv11} with no updates beyond the error-flagging outlined in Sec. \ref{subsec:flag}; the approach is summarized here for completeness.

In this approach, the complex-valued probe-induced phase and amplitude are required, and are estimated using the methods in Sec. \ref{subsec:pp} to give complex $\Delta p_1$, $\Delta p_2$, and $\Delta p_3$ for the three probe pairs CGI uses.  (This approach can be straightforwardly generalized to any number of pairs.)  For each probe pair, an additional empirical value $\delta_{n}$ is calculated:
\begin{align}
\delta_{n} &= \frac{I_{+,n} - I_{-,n}}{2} \\
&\approx \frac{\left|C[E_{0}]  + i \Delta p_n\right|^2 - \left|C[E_{0}]  - i \Delta p_n\right|^2}{2}\\
&\equiv -2 \mathbf{Re}\left(C[E_{0}]\right) \mathbf{Im}\left(\Delta p_n\right) + 2 \mathbf{Im}\left(C[E_{0}]\right) \mathbf{Re}\left(\Delta p_n\right)
\end{align}
which can be rearranged in a form amenable to linear algebra:
\begin{align}
\begin{bmatrix}
-2 \mathbf{Im}\left(\Delta p_1\right) & 2 \mathbf{Re}\left(\Delta p_1\right) \\ -2 \mathbf{Im}\left(\Delta p_2\right) & 2 \mathbf{Re}\left(\Delta p_2\right) \\ ...
\end{bmatrix}
\begin{bmatrix}
\mathbf{Re}\left(C[E_{0}]\right)  \\ \mathbf{Im}\left(C[E_{0}]\right)
\end{bmatrix} &=
\begin{bmatrix}
\delta_{1}  \\ \delta_{2} \\ ...
\end{bmatrix}  
\end{align}
A number of standard techniques can be used for the $Ax = b$ solve to compute the real and imaginary parts of $C[E_{0}]$; CGI HOWFSC calls \texttt{numpy.linalg.lstsq}, which uses an SVD-based algorithm and returns a condition number on the solve, which is used for one of the cuts described in Sec. \ref{subsec:flag}.  Note that if a probe pair was previously identified as ``bad'' (see section \ref{subsec:flag}), that probe pair is simply omitted from the calculation by removing the row and the $\delta$ when building $A$ and $b$.  This calculation is repeated, at each wavelength, for each pixel that was not previously identified as bad.

At the end, the coherent intensity estimate at each wavelength is calculated as 
\begin{align}
I_{coh} &= \left|\mathbf{Re}\left(C[E_{0}]\right)\right|^2 + \left|\mathbf{Im}\left(C[E_{0}]\right)\right|^2,
\end{align}
and the incoherent intensity is:
\begin{align}
I_{incoh} &= I_{0} - I_{coh}, \label{eq:iinc2}
\end{align}
with $I_{0}$ the unprobed image at that wavelength.

\subsection{Flagging Estimation Errors} \label{subsec:flag}

\begin{figure}
\centering
\includegraphics[width=0.3\textwidth]{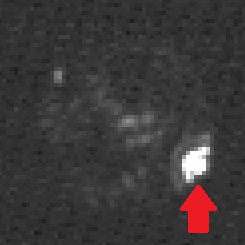}
\caption{A single badly estimated pixel caused the wavefront control to create a bright localized correction to fix it (shown with arrow)}
\label{fig:spike}
\end{figure}

\noindent Poor electric-field estimates from the wavefront-estimation step may drive the wavefront control in an unwanted direction.  Figure \ref{fig:spike} shows an example: a single pixel in the previous iteration was estimated to have an electric field approximately 1e7 times larger than physically realistic.  This was due to a combination of a floating-point type conversion and an series of noisy probe values that were estimated to be above zero at the level of floating-point error and then used in a divisor.  Image data for the next iteration showed a large localized spike as the wavefront control attempted to compensate for that pixel.  To avoid a repeat of this scenario and similar unwanted behaviors, several cuts are used to eliminate pixels from electric-field estimation at different stages of the algorithm, and each is applied per-pixel.

\begin{enumerate}
\item For each probe pair, if any of the pixels in the intensity measurements were flagged as bad, that pixel is flagged as bad for that probe pair.  Pixels may have been flagged bad on-board (from cosmic-ray masking or the fixed bad-pixel map).  The ground-in-the-loop (GITL, see Sec. \ref{subsec:gitl}) pipeline will also flag pixels as bad if they fall outside the set of pixels used in the focal-plane in the model, to avoid spending computation on pixels that will not be used.
\item For each probe pair, if the empirical estimate $I_{p}$ from Eq. \ref{eq:ip} gives $I_{p} \leq 0$, that pixel is flagged bad for that probe pair.  (Note this check is pragmatic: the $I_{p} < 0$ case needs to be excluded before taking a square root, while the $I_{p} = 0$ can give a divide-by-zero error later if allowed through.  Bad estimates from noisy data are checked later in the estimation.)
\item At the estimation level, an operator-adjustable configuration parameter sets the minimum number of good probe-pair pixels that must be present to perform the estimation.  If below this level at a pixel, the electric-field estimate is marked as bad without running any calculation.  Minimum is 2 pairs, to provide enough free parameters to solve for the electric-field components.
\item After the estimation, the condition number of the solution is checked.  If it is less than an operator-adjustable configuration parameter, the electric-field estimate is marked as bad.  This cut targets data that is not internally consistent across probe pairs.  Reference values for this parameter are based on experience from HCIT.
\item After the estimation, the incoherent intensity is checked.  If it is sufficiently negative--by a factor set by an operator-adjustable configuration parameter--the electric-field estimate is marked as bad.  This catches nonphysical overestimates of coherent intensity.  Reference values for this parameter are based on experience from HCIT.  Note that we do not do the cut right at $I_{inc} = 0$ to avoid creating a lot of false-positive bad estimates due to noise in the images.  If there truly is no incoherent signal, $\approx50\%$ of pixels would be flagged as bad by a $I_{inc} = 0$ cut.  
\end{enumerate}

The EFC implementation used for CGI HOWFSC will temporarily mask out any estimates flagged as bad, removing them from the control problem completely for the duration of the iteration.  A final defense against spurious localized estimates also comes as a side-effect of EFC-based wavefront control itself: it is not physically possible for the DMs to correct only a single pixel at a time.  Any local correction will spill its PSF wings over onto neighboring pixels, and trying to make a single bad pixel better will make neighboring pixels worse.  (This can be seen in Fig. \ref{fig:spike}.) The least-squares solver in EFC-based control will select a solution that balances these effects.

Note that all of these cuts are targeted at mitigated errors due to random, transient effects (detector noises, shot noise, cosmic rays).  Cosmic rays in particular are expected to arrive at a cadence of 5/cm$^2$/sec, and for a $153 \times 153$ subarray of 13$\mu$m square pixels used for HOWFSC (see Sec. \ref{subsec:gitl_imp}), we expect around 0.2 hits/sec in each frame.  Estimation errors due to model mismatch, leading to poor probe-phase estimates (see Sec. \ref{subsec:pp}), will only be caught by these cuts to the extent that they make the electric-field estimate non-physical.  Model mismatch issues more generally can be investigated by looking at metrics between iterations, such as comparing the complex-valued correlation between the change in the measured and modeled electric fields.

\section{CGI High-Order Wavefront Control Approach}

\subsection{Wavefront Control with EFC} \label{subsec:efc}

\noindent Once we have estimated the electric field associated with each pixel, we want to be able to use that information to control the residual coherent speckles we measured.  For CGI, we use an algorithm called electric field conjugation (EFC \mycitep{Giv07}) to do this.

EFC attempts to move the DM actuators in a way that minimizes a locally-linearized model of the electric field as a function of actuator motion.  Specifically, if we consider a vector of $N_{DM}$ DM actuator positions $\mathbf{a}$ and an electric field $E(\mathbf{a})$ over some set of focal-plane regions $S$, we can write the first-order expansion of that field about a specific DM setting $\mathbf{a_{0}}$ as:
\begin{align}
E(S; a) &= E(S; \mathbf{a_{0}}) +  \left. \dd{E(S; \mathbf{a})}{\mathbf{a}}\right|_{\mathbf{a} = \mathbf{a_{0}}} (\mathbf{a} - \mathbf{a_{0}}) + \mathcal{O}\left(\mathbf{a}^2\right) .
\end{align}
If assume $\mathbf{a}$ is small enough to neglect higher-order terms than linear, then we can approximately minimize the intensity $||E(S; \mathbf{a})||_2$ over the region $S$ by 
\begin{itemize}
\item selecting a set of $N_{pix}$ points within that region $S$, 
\item computing $E(S; \mathbf{a_{0}})$ and $\left.\dd{E(S; \mathbf{a})}{\mathbf{a}}\right|_{\mathbf{a} = \mathbf{a_{0}}}$ at each point and storing them as two matrices $\mathbf{E}$ and $\mathbf{J}$ ($N_{pix} \times 1$ and $N_{pix} \times N_{DM}$, respectively)
\item Letting $\Delta \mathbf{a} \equiv (\mathbf{a} - \mathbf{a_{0}})$ and solving for $\Delta \mathbf{a}$ to minimize $||\mathbf{E} +  \mathbf{J} \Delta \mathbf{a}||$.
\end{itemize}
This is the simplest form of EFC: $\mathbf{E}$ are the electric fields measured as part of wavefront estimation (see Sec. \ref{subsec:est}), $\mathbf{J}$ (the ``Jacobian'') is calculated from a model (see Sec. \ref{subsec:jac}), and the solve is a least-squares ``$Ax=b$'' solve that can be implemented with any number of standard linear algebra tools.  (Technically the result can be written in closed-form as
\begin{align}
\Delta \mathbf{a} &= - (\mathbf{J}^T \mathbf{J})^{-1} \mathbf{J}^T \mathbf{E}, \label{eq:orig_da}
\end{align}
but in practice there are more computationally-efficient ways of solving for $\Delta \mathbf{a}$ than building the inverse of a very large matrix.)

As this step only solves a linearized approximation to the problem, it is necessary to collect additional data and iterate in order to achieve deep contrast.  For CGI, we augment this basic approach with a number of enhancements:
\begin{itemize}
\item We split $\mathbf{J}$ and $\mathbf{E}$ into real and imaginary parts, and use $2N_{pix} \times N_{DM}$ and $2N_{pix} \times 1$ real-valued matrices rather than $N_{pix} \times N_{DM}$ and $N_{pix} \times 1$ complex-valued matrices.  This matches the implementation approach in \mycitet{Giv07} and saves computational effort in calculating imaginary components that should be identically zero.  Using real-valued matrices ensures $\Delta \mathbf{a}$ is automatically real-valued.
\item We measure the electric field at several wavelengths separately at each iteration using the pairwise probing process.  However, we solve for all wavelengths at once in the correction step: $\mathbf{E}$ is augmented to cover electric-field estimates at several wavelengths ($N_{pix}$ becomes the total number of electric-field estimates across all wavelengths), and the model calculations for $\mathbf{J}$ are extended to cover multiple wavelengths.  See \mycitet{Giv07} for a full matrix representation.  Prior modeling (e.g. \mycitet{Zho16}) has shown the importance of chromatic control to reaching deep contrasts for CGI.
\item We use 2 deformable mirrors in the vector $\mathbf{a}$, and $N_{DM}$ becomes the total number of actuators across both DMs
\item We apply a weighting matrix to each electric-field estimate ($W_{E}$) and to each DM actuator ($W_{DM}$) to allow regions of the focal plane or the DMs to be emphasized, ignored, or linked together.  See Sec. \ref{subsec:ppw} and Sec. \ref{subsec:paw} respectively for the details of the creation of each.
\item We use a scalar regularization parameter so that the least-squares solve applies weight to the magnitude of the delta-DM setting being applied.  Practically speaking, this modifies Eq. \ref{eq:orig_da} to:
\begin{align}
\Delta \mathbf{a} &= - (\mathbf{J}^T \mathbf{J} + \lambda \mathbf{I})^{-1} \mathbf{J}^T \mathbf{E}, \label{eq:reg_da}
\end{align}
with $\mathbf{I}$ an $N_{DM} \times N_{DM}$ identity matrix.  
\end{itemize}
The final ``$Ax=b$'' problem we solve for CGI is written as:
\begin{align}
\left(W_{DM}^T J^T W_{E}^T W_{E} J W_{DM} + \lambda \mathbf{I}\right) \Delta \mathbf{a} &= -W_{DM}^T J^T W_{E}^T W_{E} \mathbf{E}. \label{eq:axb}
\end{align}
Three different methods of performing this solve have been implemented--and cross-checked--in CGI ground software (GSW): Cholesky decomposition, QR decomposition, and an iterative preconditioned conjugate-gradient solver.  All produce the same results to floating-point precision for non-pathological inputs; in practice Cholesky is our default as the ``$A$'' matrix is square and normal.

\subsubsection{Per-pixel weighting} \label{subsec:ppw}

\noindent The $W_{E}$ weighting matrix is, in principle, a $2N_{pix} \times 2N_{pix}$ matrix that applies weights to electric-field elements.  In practice, we made the decision that electric-field elements will never be coupled in the weighting matrix, and so $W_{E}$ is diagonal.  $W_{E}$ is built from four sources:
\begin{enumerate}
\item The entire $W_{E}$ is populated from the weighting matrix in the control strategy (see Sec. \ref{subsec:cs}).  The same weight is applied to real and imaginary parts for a given electric-field component.  Ones in the matrix indicate regular weighting.
\item The HOWFSC software applies an internal correction per wavelength ($\times \lambda_{center}/\lambda$) to compensate for the fact the Jacobian actuator pokes are computed in radians, but the actual settings are applied in nm.  This is handled automatically in software and does not need to be built into the control strategy.
\item Any fixed bad-pixels are given a weight of 0.  This information is extracted from the control strategy.
\item Any electric-field elements marked bad in a given iteration are also given a weight of 0.
\end{enumerate}
All of $W_{E}$ except the per-iteration bad-electric-fields can be precalculated (see Sec. \ref{subsec:jac} on Jacobian products); these are incorporated into a low-rank update to $J^T W_{E}^T W_{E} J$ during the wavefront correction step rather than being added to $W_{E}$ directly.

\subsubsection{Per-actuator weighting} \label{subsec:paw}

\noindent The $W_{DM}$ weighting matrix is a $N_{DM} \times N_{DM}$ matrix that applies weights to individual actuators of DMs.  Unlike the $W_{E}$ matrix, this matrix is not generated from control strategy contents.  Rather, it starts with an identity matrix and is assembled during the control step to capture two DM-actuator behaviors used in HOWFSC: ``freezing'' and ``tying''.
\begin{itemize}
\item Frozen actuators will not be moved by wavefront control.  An actuator is frozen by zeroing the associated diagonal element in $W_{DM}$.
\item Tied actuators will be moved by wavefront control the same amount.  (This does not require they have the same voltage, only that they move together.)  An actuator is tied by making the following modification to the subset of rows/columns of $W_{DM}$ associated with the tied actuators:
\begin{align}
\mathrm{M\, tied\,actuators}\left\{\begin{bmatrix}
1 & 0 & \hdots & 0 \\ 
0 & 1 &   & 0 \\ 
\vdots &   & & \vdots \\
0 & 0 & \hdots & 1 \\
\end{bmatrix}\right. & \Rightarrow
\begin{bmatrix}
\frac{1}{M} & \frac{1}{M} & \hdots & \frac{1}{M} \\ 
\frac{1}{M} & \frac{1}{M} &   & \frac{1}{M} \\ 
\vdots &   & & \vdots \\
\frac{1}{M} & \frac{1}{M} & \hdots & \frac{1}{M} \\
\end{bmatrix}
\end{align}
\end{itemize}
While the tied-actuator constraints make the matrix no longer diagonal, the entire $W_{DM}$ matrix is well-suited to sparse-matrix formulation and is implemented this way for CGI HOWFSC.

Freeze and tie constraints come from the following three sources:
\begin{enumerate}
\item Each DM has a \textit{tiemap} defined for it.  For CGI, these tiemaps are $48 \times 48$ matrices, where a 0 indicates no tie, and a unique nonzero integer $1 \rightarrow N$ for each of the $N$ tie groups, which are regions of the DM outside the dimension of the pupil which were electrically connected to a single voltage channel to save mass and power in the DM driver boards.  $-1$ are used to indicate actuators which are nonfunctional.  Dead actuators are not included in tie groups, even if they are known to be electrically-tied, as in the corners or center of the DMs.  The tiemaps for the CGI DMs are shown in Fig. \ref{fig:tiemaps}.

\begin{figure}
\centering
\includegraphics[width=1\textwidth]{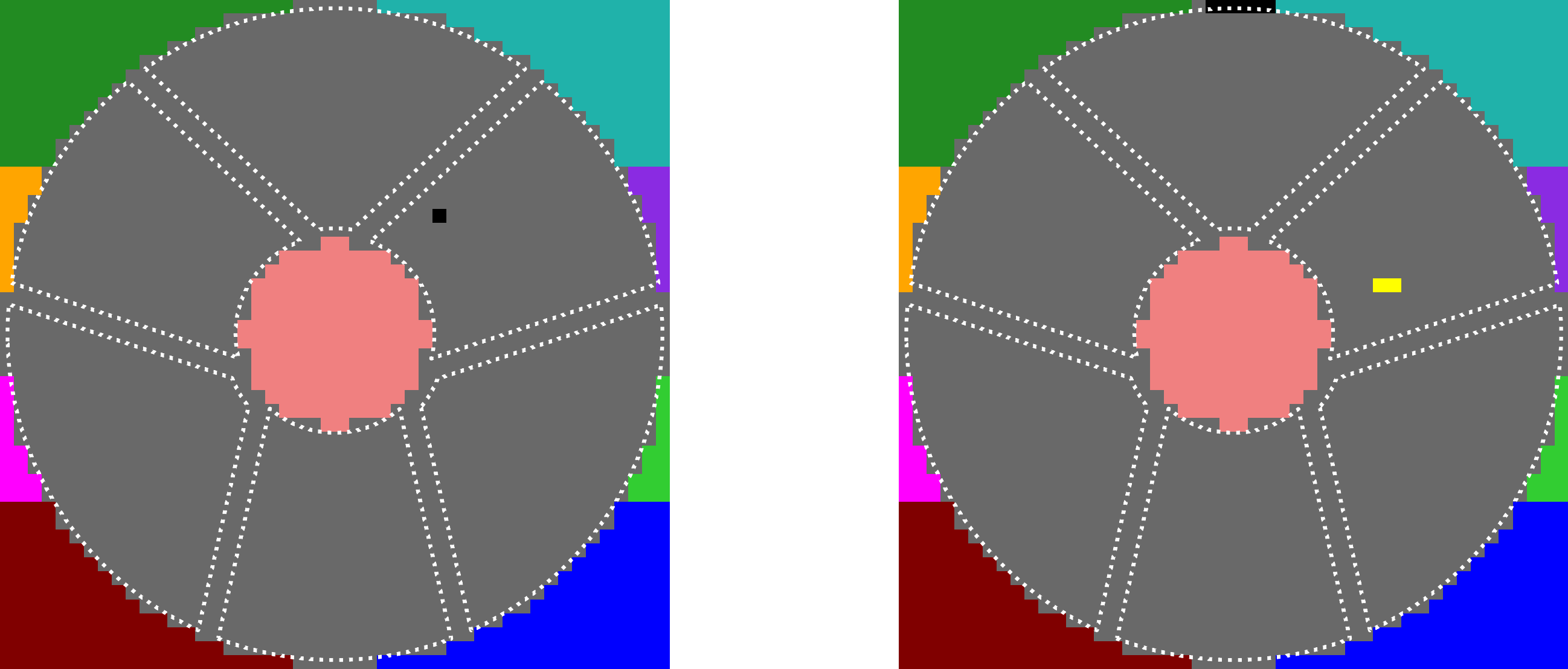}
\caption{\textit{Left:} Tiemap for DM1, post-TVAC. \textit{Right:}  Tiemap for DM2, post-TVAC.  Both tiemaps have standard actuators (not dead or tied) in gray, dead actuators in black, and separate colors for each individual group of tied actuators.  All tie groups are intentionally tied together, except for the pair of actuators in yellow at 3 o'clock on DM2, which were discovered to be tied together during assembly-level hardware characterization.  The Roman pupil, as seen on each DM, is overlaid.} \label{fig:tiemaps}
\end{figure}

Tiemaps are updated only rarely, in the event of a change in the electrical-connectivity topology detected during other calibration activities, and never changed during HOWFSC.  All members of each individual tie groups are internally tied to each other in $W_{DM}$.  All dead actuators are frozen in $W_{DM}$.
\item The CGI DMs are limited in their motion due the finite range (0-100V) made available to the DMs.  In any given iteration, if the initial DM setting is railed high or low, that actuator is frozen for that iteration.  While this does in principle slightly limit the control authority available, it also prevents the iteration from trying to send an out-of-range command that would have been clipped regardless.
\item The CGI DMs also have ``neighbor rules'', limitations on the interactuator stroke differential in the lateral and diagonal directions to prevent an excess of strain from plastically deforming the epoxy holding the DM facesheet.  These limit the voltage between laterally-adjacent actuators to 50V and diagonally-adjacent actuators to 75V.  In any given iteration, if two actuators are at a neighbor-rule limit, they are tied together for that iteration.   As with voltage caps, this trades a small amount of control authority for a guarantee that the least-squares solution will not create a neighbor-rule violation.  (There is no safety issue--there are two separate routines, in HOWFSC GITL on the ground and in CGI flight software (FSW), which both detect and correct neighbor-rule violations before anything is applied--but this would cause the iteration to be less effective than expected, and can in principle cause control to stall if the least-squares solution preferred by EFC is an otherwise-illegal move.)
\end{enumerate}

\subsection{The HOWFSC Optical Model} \label{subsec:model}

\noindent CGI HOWFSC uses a diffractive optical propagation model with three primary applications during GITL:
\begin{itemize}
\item calculate probe phases (wavefront estimation, Sec. \ref{subsec:pp})
\item calculate Jacobians (wavefront correction, Sec. \ref{subsec:jac})
\item calculate contrast estimates for the next iteration (operator monitoring, camera-settings calculation)
\end{itemize}
Other uses are offline, but just as important: 
\begin{itemize}
\item creating model-based DM ``seeds''--DM settings built by running the wavefront control model against itself by simulating data frames from the instrument--to initialize wavefront control iterations,
\item creating simulated data frames to test control strategies against, 
\item scaling DM probe patterns,
\item debugging and regression testing when making updates to the modeling software.
\end{itemize}

\begin{figure}
\centering
\includegraphics[width=1\textwidth]{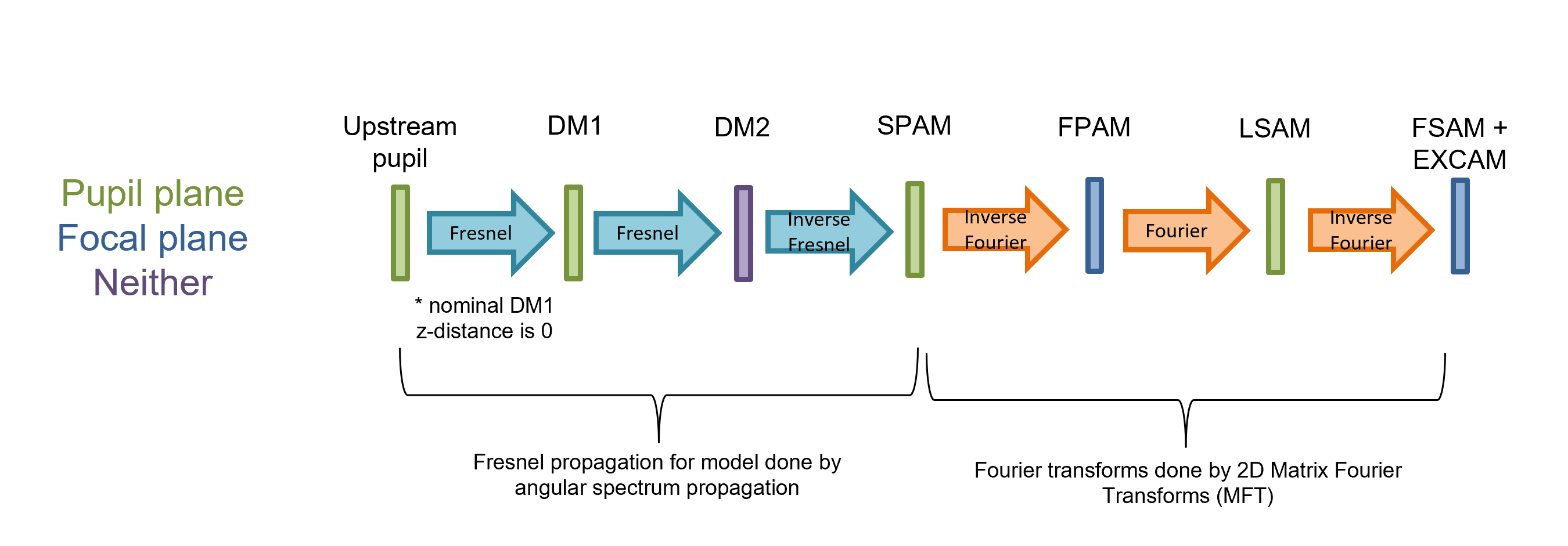}
\caption{A block diagram of propagation through a HOWFSC optical model, per wavelength.  All planes other than the initial and final planes correspond to mechanisms within CGI: 2 deformable mirrors (DM1 and DM2), and 4 Precision Alignment Mechanisms (PAMs), 2-axis X-Y stages which allow masks, lenses, filters, and other optical elements to be moved in and out of the beam.  2 PAMs are in pupil planes: the Shaped Pupil Alignment Mechanism (SPAM) and the Lyot Stop Alignment Mechanism (LSAM).  2 PAMs are in focal planes: the Focal-Plane Alignment Mechanism (FPAM) and the Field Stop Alignment Mechanism (FSAM).  All planes are represented at the resolutions and orientations as seen on EXCAM, except at FPAM, where higher resolution is used for the focal-plane mask.} \label{fig:hmod}
\end{figure}

The CGI HOWFSC model is ``compact'': it only includes planes with masks or DMs, and projects all wavefront aberrations derived from phase retrieval into two planes, one upstream of the focal-plane mask and one downstream.  A separate full high-fidelity model with all optical surfaces is used for detailed performance modeling, and for structural-thermal-optical (STOP) modeling with realistic structural and thermal deformations converted to optical aberrations as a function of time; see \mycitet{Kri23} for further details.  The layout of the compact model is shown in Fig. \ref{fig:hmod}; Fresnel propagations are done with angular spectrum propagation \mycitep{Goo96} and propagations to mask planes and the final focal plane are done with 2D matrix Fourier transforms (MFTs \mycitep{Sou07}).

A HOWFSC model consists of two parts: a HOWFSC optical model definition file, and a software framework to build a propagation-model object instance from that data file.  

As a design principle, the model definition file must contain enough data elements to define the entire optical model.  To satisfy this, the model definition file is populated with three categories of data:
\begin{enumerate}
\item For each wavelength, the complex-valued properties of each coronagraphic mask must be specified, along with any tip/tilt or residual wavefronts from system optics upstream and downstream of the focal plane mask.  Each also has a boolean array indicating which pixels are included in the wavefront estimation and control following the model.  These are used to build one monochromatic propagation object per wavelength.  Note that for the CGI control model, we use one wavelength per engineering subfilter (e.g. 3 subfilters across CGI Band 1, which is a 10\% band centered at 575nm), but the HOWFSC model structure does not require or hard-code any particular number of wavelengths nor any particular spectral sampling.
\item A single pair of DM settings which correspond to the DM commands when the upstream wavefront was measured.  These are common to all wavelengths, and the upstream settings in the model should all have been collected at the same wavelength.
\item For each DM, we define all the values to map a $48\times48$ command into a wavefront with dimensions consistent with pupil imaging on EXCAM.  (It is not a requirement that the focal-plane and pupil-plane model components match camera sampling, but we do it whenever possible to permit direct comparisons of model vs. measured.)  These include actuator pitch and influence function, and translation, rotation, scaling, and parity (i.e. mirror-image the DM map or not) values necessary to place that grid of influence functions in pupil space.  We also capture electrical properties of the DM and its electronics, including voltage ranges, gainmaps, tiemaps, and maps of crosstalk between actuators.  These are all independent of wavelength.
\end{enumerate}
The software framework ingests this file and creates an instance of a model object with methods to permit propagation with different DM settings and wavelengths, as well as propagation of derivatives for Jacobian calculation.  GITL and other software pieces may call these methods to use the model.

For practical reasons, the model definition file must be data, not code.   Keeping the models outside the software permits them to be updated at timescales faster than the release schedule for CGI ground software if need be, and permits them to be viewed and potentially updated by personnel who are subject-matter-experts but not software engineers.  To this end, we also required that model definition files be able to be created, read, and modified by humans as well as computers.  The HOWFSC models are written in YAML, a data-serialization language designed for human readability.  Other choices of language (e.g. JSON, XML) are possible and likely could also have been implemented for this application as well.  Several ground software functions were created to aid in updating a HOWFSC model YAML file, though manual updates are also valid.

For reliability, model definition files must be able to be validated before use.  On CGI, this was done by creating a written specification for the YAML files that store HOWFSC models, and writing a validator against this specification which was delivered with the CGI ground software.  This validator is exposed and may be called by personnel outside of GITL when preparing material; it is also called automatically in GITL operation during input validation by the software framework.

We note that it is not sufficient that a HOWFSC optical model be structurally correct; the contents must also correctly represent the instrument as-built to a high precision.  In our 2014-2017 Milestone testing, we found that model-mismatch was the primary reason wavefront control slows or stops at moderate contrast.  Describing all of the calibration data collection and processing activities used to populate the optical model is beyond the scope of this paper; the reader is referred to \mycitet{Rig23} for CGI calibration-algorithm descriptions and \mycitet{Zho18} for sensitivity analysis to model-mismatch errors.

\subsection{Jacobian Calculation} \label{subsec:jac}

\noindent In order to set up the solve in Eq. \ref{eq:axb}, we need to populate the Jacobian $J$.  Let us treat the coronagraph as two linear operators $\mathcal{C}_{up}$ and $\mathcal{C}_{down}$, one representing propagation from an input phase $A$ to a DM (upstream) and one for propagation from the DM to the final focal plane (downstream), and consider with one actuator $\alpha$ with influence function $\phi$:
\begin{align}
E(0) &= \mathcal{C}_{down}\left(\mathcal{C}_{up}(A)\right), \\
E(\alpha) &= \mathcal{C}_{down}\left(\mathcal{C}_{up}(A) e^{i \phi \alpha}\right),
\end{align}
then the derivative with respect to that actuator motion at $\alpha = 0$ is:
\begin{align}
\left.\dd{E(\alpha)}{\alpha}\right|_{\alpha = 0} &= \mathcal{C}_{down}\left(i \phi \mathcal{C}_{up}(A)\right).
\end{align}
That is, we can multiply by $i \phi$ at the DM plane in the model to compute the Jacobian contents for a single actuator, and repeat over all actuators.  Note that with two DMs, $\mathcal{C}_{up}$ and $\mathcal{C}_{down}$ will not be the same for both DMs.

Each of these calculations is independent of each other, and so CGI enables multiprocessing to split this calculation over many cores.  CGI gains further efficiency by precalculating and reusing $C_{up}(A)$, which is the same for each DM, and by implementing the fast Jacobian calculation from FALCO \mycitep{Rig18}: as the influence function $\phi$ is localized at the DM plane and modeled as zero outside that local region, only a subregion needs to be propagated through parts of the model.

Finally, CGI slightly modifies the Jacobian to minimize intensity relative to PSF peak (``normalized intensity'', $|E/E_{pk}|^2$) rather than just intensity $|E|^2$, as described in \mycitep{Llo22}:
\begin{align}
J_{pk} &= \frac{\dd{E(\alpha)}{\alpha}}{E_{pk}} - \frac{E(\alpha)}{E_{pk}^2} \dd{E_{pk}}{\alpha}
\end{align}
This ``peak Jacobian'' ensures the PSF remains sharp for off-axis sources. 

\begin{figure}
\centering
\includegraphics[width=1\textwidth]{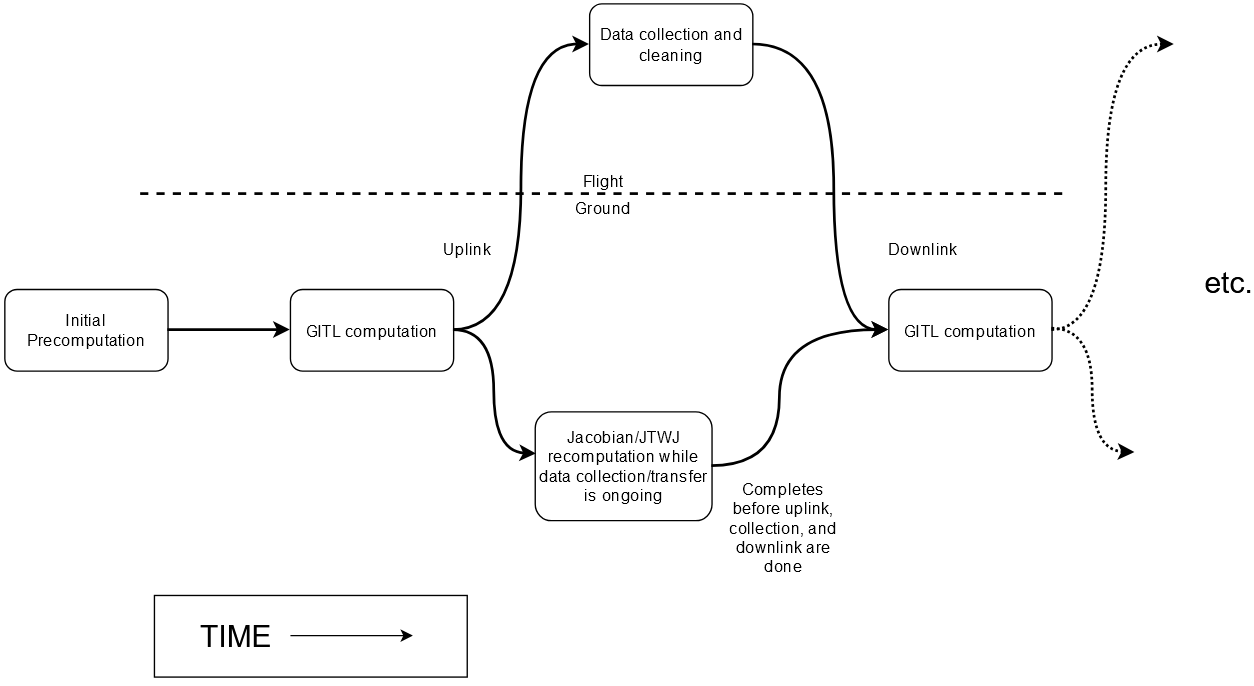}
\caption{Data flow for early-Jacobian calculations} \label{fig:earjac}
\end{figure}

Operationally, it turns out to be necessary to recalculate the Jacobian every iteration, particularly after large DM motions.  We accomplish this with the ``early Jacobian'' approach: all of the information necessary to compute a Jacobian for iteration $I+1$ is known once the DM settings for iteration $I$ are available, and so the calculation can be kicked off in parallel while iteration $I+1$ is collecting data for wavefront estimation.  (See Fig. \ref{fig:earjac}.)  As long as the Jacobian calculation completes before all of the iteration $I+1$ data has been downlinked and made available, it adds zero downtime to operations to compute a Jacobian every iteration.

When computing the Jacobian, we also compute $J^T W_{E}^T W_{E} J$ and store it to save computation time during the control step.  This is currently the only ``Jacobian product'' being produced by CGI.

For completeness, we note that since the development of the CGI HOWFSC code base, advances in algorithmic differentiation have provided paths to do calculations without explicitly computing a Jacobian (see e.g. \mycitet{Wil21}).  We have elected not to pursue these at this time for CGI HOWFSC, not because there is anything wrong with them, but because the Jacobian calculations and operations already function and meet requirements, and we are focusing our team's efforts on things that are still incomplete.  It may be revisited in the future with an eye toward supporting HWO (Habitable Worlds Observatory) technology advancement if time and funding permit.

\section{CGI High-Order Wavefront Control Strategies}

\subsection{Control Strategies} \label{subsec:cs}

\noindent While most of HOWFSC operations apply an identical mathematical framework to incoming HOWFSC GITL frames regardless of circumstances, there are a handful of parameters which can be varied per iteration to adjust the wavefront control performance.   These are the only knobs available to operators/subject-matter experts to incorporate prior lessons learned about how to successfully create a region of deeply-suppressed starlight for a particular coronagraph configuration.  CGI uses a particular structure for capturing and managing all of these called a ``control strategy'' \mycitep{Mar17}.  

As with HOWFSC optical model definition files (Sec. \ref{subsec:model}), control strategies must contain enough data elements to select all HOWFSC inputs that vary per iteration in a hands-off manner.  For CGI, we eventually settled on seven pieces of data in the control strategy.   Three items are used to prepare the wavefront-control solution with EFC:
\begin{itemize}
\item Regularization parameter ($\lambda$ in Eq. \ref{eq:axb}).  See Sec. \ref{subsec:bb} for a detailed discussion on the use of regularization and conversion from $\lambda$ to $\beta$; we used values in the range $\beta = -2$ through $\beta = -6$ during CGI TVAC.
\item Amplitude weighting matrix per pixel.  This is $N_{\lambda}$ 153x153 arrays for each of the $N_{\lambda}$ wavelengths used in the control.  (See Sec. \ref{subsec:gitl_imp} for the origin of the array dimensions.)  Each pixel contains a per-pixel weight to be applied to the control: unweighted pixels should have weight = 1, and weighting a pixel by $Y$ implies that control will value the reduction in intensity at that pixel by $Y^2$.  This data is applied to $W_{E}$ as described in Sec. \ref{subsec:ppw}.  As a practical matter, the arrays are stored in a file and the control strategy contains a pointer to that file.  CGI used weight = 1 for all pixels and wavelengths during CGI TVAC.
\item EXCAM bad pixel map.  This is a 1024x1024 boolean array sized to the cleaned-frame region, which indicates which pixels on the EXCAM detector have been flagged as bad by a separate calibration activity.  Telemetry channels in a GITL ancillary packet are used to crop that bad-pixel map to the 153x153 array.  Bad pixels are set to weight = 0 in $W_{E}$.  As a practical matter, this array is stored in a file and the control strategy contains a pointer to that file.  CGI did not have any bad pixels on EXCAM during CGI TVAC.
\end{itemize}

Another four items are not used in EFC, but are used to prepare DM settings and CGI FSW parameters for upload:
\begin{itemize}
\item Multiplicative gain to apply to change in DM setting (1 = use calculated $\mathbf{\Delta_{a}}$, 0.5 = use half calculated $\mathbf{\Delta_{a}}$, etc.).  This is a scalar multiplier on the DM setting change resulting from EFC, and is used to prepare the absolute DM1 and DM2 settings to be uploaded for the next iteration.  Values from 0.5 to 1.0 were used in CGI TVAC.
\item Mean probe intensity across dark hole.  We have generally found it advantageous to drop probe intensity at deeper contrasts rather than leaving it fixed through iterations, to avoid swamping the signal with probe-induced shot noise.  However, probe intensity which is too low can make the modulation signal difficult for wavefront estimation to estimate.   This value is converted into a scalar multiplier to be uploaded as an FSW parameter; it is used during HOWFSC data collection via the sequence engine.  Values in the range $10^{-5}\times$ to $10^{-7}\times$ the PSF peak intensity were used in CGI TVAC.
\item Target SNR for probed and unprobed images, tabulated separately.  These are not used as part of estimation and control, but are used by the engineering exposure time calculator to prepare EXCAM settings for the next iteration: longer exposures, or more of them, or both, are needed when going to fainter contrasts to reach the same SNR.  Camera settings are uploaded as FSW parameters and are used during HOWFSC data collection via the sequence engine.  HOWFSC used a consistent unprobed SNR of 5 and a probed SNR of 7 throughout CGI TVAC.
\end{itemize}

In order to enable hands-off operation of HOWFSC, control strategies must contain enough information to complete the HOWFSC calculation for any set of valid inputs without further input from an operator.  For CGI, we settled on two pieces of information as sufficient inputs: number of upcoming iteration and contrast at current iteration.  By convention, the very first iteration of HOWFSC (run from parameters uploaded before starting) is iteration 0, and the first upcoming iteration that uses the control strategy is iteration 1.  Both of these have a fixed lower bound (contrast = 0, iteration = 1) but their upper end is technically unbounded; iteration number takes on discrete values while contrast is continuous.

\begin{figure}
\centering
\includegraphics[width=1\textwidth]{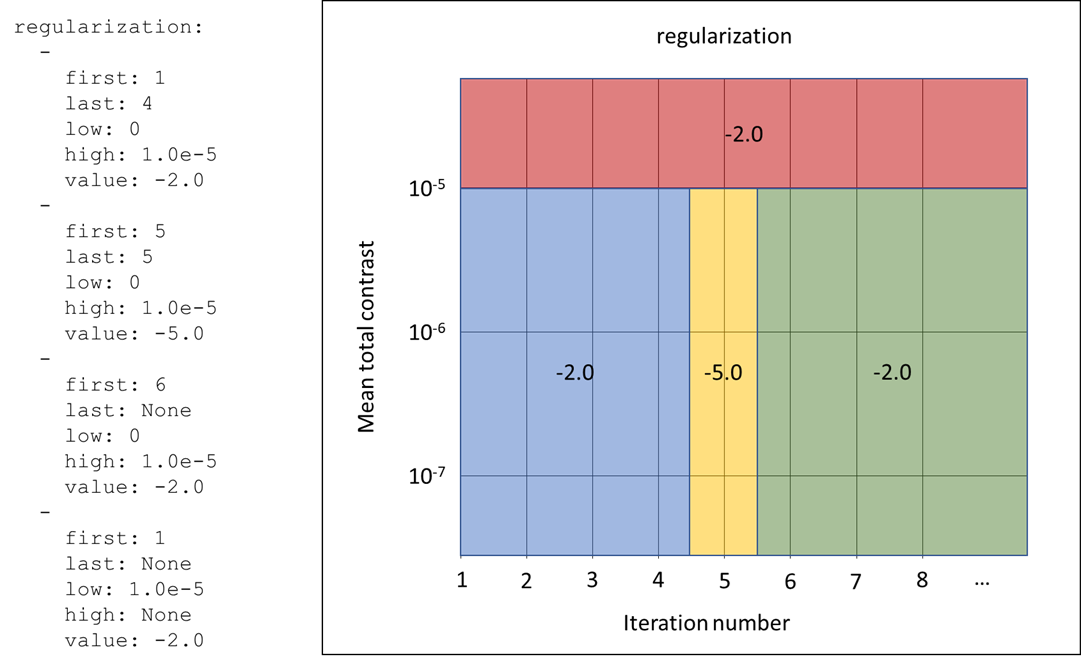}
\caption{Example of a piece of a control strategy file.  This example shows a single beta bump (see Sec. \ref{subsec:bb}) that is not applied if the contrast is very high. \textit{Left.} Excerpt from sample YAML file giving a list of regions and their associated regularization parameter.  \textit{Right.}  Diagram showing how these regions tile the space of iteration number and contrast.} \label{fig:cstrat}
\end{figure}

We implemented the control strategy as 6 custom look-up tables, one for each piece of data except the bad-pixel map (which was fixed for all iterations and contrasts).  Each custom look-up table was implemented by storing a list of regions, with five values for each region: the four boundaries of the rectangle (including an ``unbounded'' designator if necessary) and the value.  The list of regions in a look-up table must fully tile the semi-unbounded contrast/iteration space, including unbounded edges.  Each region is assigned a single value (a scalar or a string with a filename), so any valid contrast/iteration input can have its output pulled from value associated with the region that input falls into.  An example of part of a control strategy is shown in Fig. \ref{fig:cstrat}.

As with the HOWFSC model definition files, control strategies must be data, not code, to permit them to be updated outside of the slow cadence of software release.  They were implemented in YAML, chosen for the same reasons listed for the HOWFSC optical model in Sec. \ref{subsec:model}.  Also like the HOWFSC optical model, control strategy files have a written specification and a validator against that specification which can be called standalone during file preparation and is also invoked automatically during the input validation step in GITL.

\subsection{Beta bumping} \label{subsec:bb}

\noindent Of all the elements in the control strategy, regularization is one of the most important parameters to change in order to get to a high contrast.  To see this, we define $\lambda = s_{max}^2 10^{\beta}$, with $s_{max}$ the largest singular value of $W_{E} J W_{DM}$.  (During the CGI campaign, we used values for $\beta$ ranging from $-2$ to $-6$.)  This very specific definition allows for a useful decomposition of the left side of Eq. \ref{eq:axb} via singular value decomposition \mycitep{Seo17}:
\begin{align}
W_{E} J W_{DM} & \equiv U S V^T, \,\mathrm{with}\,S\,\mathrm{a\,diagonal\,matrix\,of\,singular\,values} \label{eq:USVT} \\
W_{DM}^T J^T W_{E}^T W_{E} J W_{DM} &= V S^T S V^T
\end{align}
and since $U$ and $V$ are unitary, and the identity matrix is commutative:
\begin{align}
W_{DM}^T J^T W_{E}^T W_{E} J W_{DM} + \lambda \mathbf{I} &= V S^T S V^T + s_{max}^2 10^{\beta} \mathbf{I} \\
&= V S^T S V^T + s_{max}^2 10^{\beta} \mathbf{I} V V^T = V \left(S^T S + s_{max}^2 10^{\beta}\right) V^T.
\end{align}
$S^T S$ is also diagonal, and ordered from largest value to smallest: $[s_{max}^2, \hdots, s_{min}^2]$; the act of regularizing in this manner effectively reweights singular values to $[s_{max}^2 (1 + 10^{\beta}), \hdots, s_{min}^2 + s_{max}^2 10^{\beta}]$.  

To see why this is useful, consider the exact solution to Eq. \ref{eq:axb} with and without regularization.  (Note: we do not do this calculation in practice for HOWFSC, as both computing an SVD and computing an inverse are much more computationally expensive than other options available, but it is useful to examine for understanding.)
\begin{align}
\left(W_{DM}^T J^T W_{E}^T W_{E} J W_{DM} + \lambda \mathbf{I}\right) \Delta \mathbf{a} &= -W_{DM}^T J^T W_{E}^T W_{E} \mathbf{E} \nonumber
\end{align}
For $\lambda = 0$:
\begin{align}
\Rightarrow \Delta \mathbf{a} &= -\left(W_{DM}^T J^T W_{E}^T W_{E} J W_{DM}\right)^{-1} W_{DM}^T J^T W_{E}^T W_{E} \mathbf{E} \\
&= -\left(V \mathrm{diag}[s_{max}^2, \hdots, s_{min}^2] V^T\right)^{-1} V S U^T W_{E} \mathbf{E} \\
&= -V \mathrm{diag}[s_{max}^2, \hdots, s_{min}^2]^{-1} V^T V S U^T W_{E} \mathbf{E} \\
&= -V \mathrm{diag}\left[\frac{1}{s_{max}}, \hdots, \frac{1}{s_{min}}\right] U^T W_{E} \mathbf{E}
\end{align}
For $\lambda \neq 0$:
\begin{align}
\Rightarrow \Delta \mathbf{a} &= -\left(W_{DM}^T J^T W_{E}^T W_{E} J W_{DM} + \lambda \mathbf{I}\right)^{-1} W_{DM}^T J^T W_{E}^T W_{E} \mathbf{E} \\
&= -\left(V \mathrm{diag}[s_{max}^2 (1 + 10^{\beta}), \hdots, s_{min}^2 + s_{max}^2 10^{\beta}] V^T\right)^{-1} V S U^T W_{E} \mathbf{E} \\
&= -V \mathrm{diag}[s_{max}^2 (1 + 10^{\beta}), \hdots, s_{min}^2 + s_{max}^2 10^{\beta}]^{-1} V^T V S U^T W_{E} \mathbf{E}\\
&= -V \mathrm{diag}\left[\frac{1}{s_{max}(1 + 10^{\beta})}, \hdots, \frac{1}{s_{min} + \frac{s_{max}^2}{s_{min}} 10^{\beta}}\right] U^T W_{E} \mathbf{E}
\end{align}
with a ratio between the two of:
\begin{align}
\left[\frac{1}{1 + 10^{\beta}}, \hdots, \frac{1}{1 + \left(\frac{s_{max}^2}{s_{min}^2}\right)10^{\beta}}\right]
\end{align}
Effectively, this regularization scheme reweights all of the SVD modes, penalizing ones where $\frac{s^2}{s_{max}^2} < 10^{\beta}$ while minimally affecting modes $\frac{s^2}{s_{max}^2} > 10^{\beta}$.  A large $\beta$ effectively penalizes ``hard modes'', SVD modes that use a lot of DM stroke to create a small change in electric field, in favor of ``easy modes''.  Pupil-plane phase errors will generally show up as easy modes, while amplitude errors and chromatic errors will require harder modes to correct.

To get very deep contrasts, we need to be able to access hard modes with correction.  However, using only aggressive values of $\beta$ will result in a suboptimal contrast floor due to nonlinearity and calibration error; see discussions in \mycitet{Mar17} and \mycitet{Sid17}.  Instead, we schedule our $\beta$ values as a function of iteration to alternate between ones that can access both easy and hard modes, and ones that can access easy modes only and clean up residuals from the aggressive correction.  Empirically, this is found to ratchet down the contrast in the end, though it will usually spike briefly after an aggressive correction.  These spikes can be seen in contrast vs. iteration curves in Secs. \ref{subsec:nfov} and \ref{subsec:wfov}.  See \mycitet{Rua22} for examples of beta-bumping applied over longer timescales (though not for CGI applications).

We note that using a single scalar multiplier $\lambda$ on an identity matrix is not the only way of applying regularization; this is a special case (``ridge regression'') of a generalized class of regularizations (``Tikhonov regularization'')\mycitep{Gol13} which permits a wider variety of matrices.  This expanded control space was not investigated by the CGI team, though others may find it useful to know the additional degree of freedom exists.

\subsection{Other HOWFSC Operational Considerations}

\noindent On CGI, the low-order wavefront sensing and control (LOWFSC) and HOWFSC systems are conceptually disjoint: LOWFSC ``locks in'' the wavefront against jitter and drift, while HOWFSC makes fine corrections to that stabilized wavefront.  However, they interact in two important ways:
\begin{itemize}
\item LOWFSC has three loops, a fast line-of-sight loop controlled by the Fast Steering Mirror (FSM) with 20Hz bandwidth, and two slower loops, a focus loop with the fine stage of the Focus Control Mechanism, and a ``Zernike'' loop on DM1 which uses the DM surface to create patterns of astigmatism, coma, trefoil, and primary spherical aberration.  Since both LOWFSC and HOWFSC want to use DM1, we need a way to integrate commands from both.
\item HOWFSC patterns from EFC may contain components that look like tip, tilt, focus, etc. but are part of the desired correction.  If we apply these settings while LOWFSC loops are locked, we want to make sure that LOWFSC does not try to undo part of the correction just applied.
\end{itemize}

To solve both of these issues, CGI implemented a ``DM combiner'', which is executed when a DM command (absolute or relative) is applied from any source.  It merges the command with the existing setting, including any neighbor rule or voltage limit considerations, and updates the setpoints of all three LOWFSC loops so that the loops do not try to correct for anything intentionally induced by a DM command.  This DM-combiner approach was first developed for technology milestone testing, and was successfully demonstrated in 2017 to run LOWFSC and HOWFSC together \mycitep{Shi17a}.  For CGI, it was subsequently reimplemented in flight software.

\section{The Ground-in-the-Loop (GITL) Approach to HOWFSC} \label{subsec:gitl}

\subsection{Definition of GITL approach}

\noindent On CGI, HOWFSC is implemented in a ground-in-the-loop (GITL) manner: image data is sent to the ground for processing, and DM settings and EXCAM settings (EM gain, exposure time per frame, number of frames) are uplinked for the next iteration.  See Tables \ref{tab:gitl_down} and \ref{tab:gitl_up} for a listing of all data passed in either direction during GITL; this data flow is designed so that the ground never has to guess or assume any system properties during data collection.

\begin{figure}
\centering
\includegraphics[width=1\textwidth]{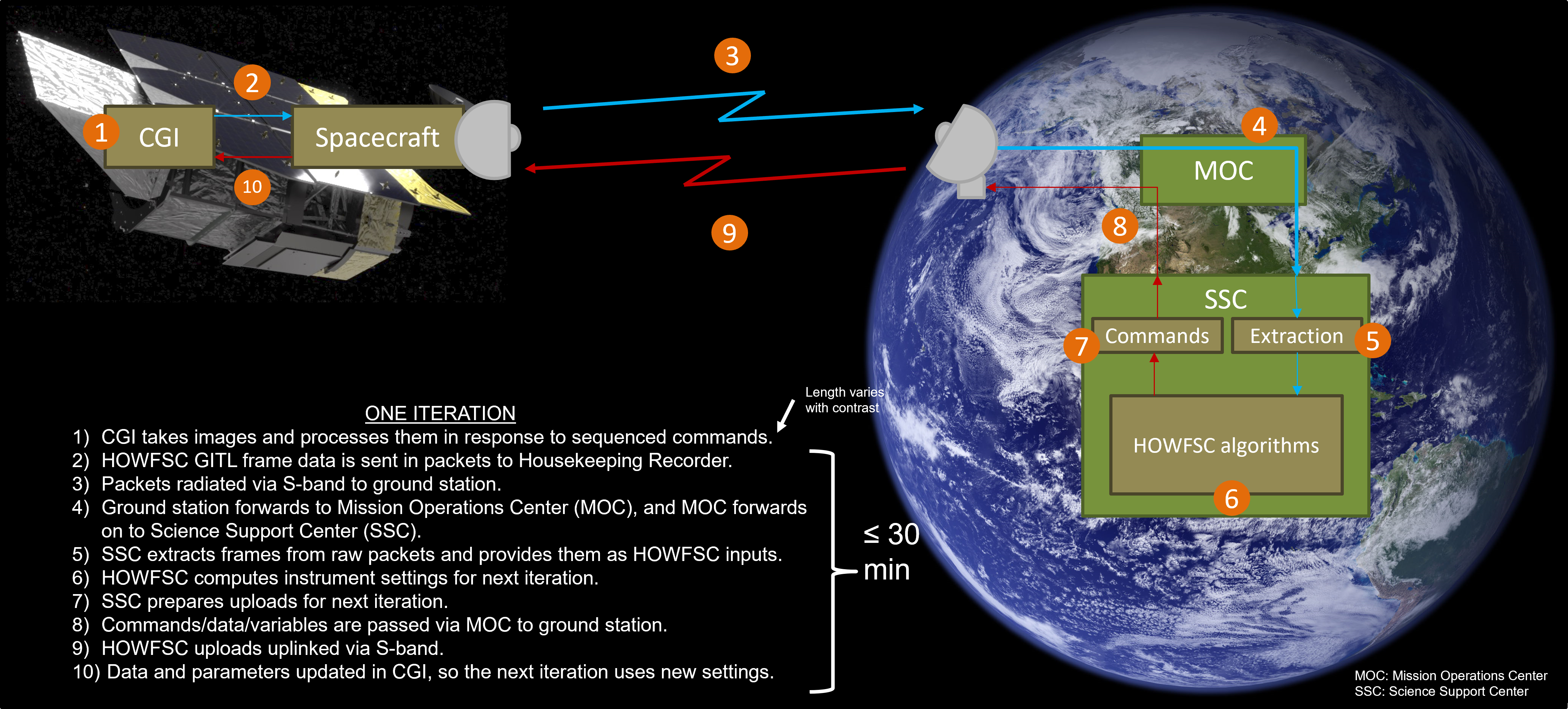}
\caption{Flow of HOWFSC data during ground-in-the-loop (GITL) operations} \label{fig:gitl_overview}
\end{figure}

This entire loop is run in a purely hands-off manner: no operator is required to provide input or run any commands at any time.  Figure \ref{fig:gitl_overview} shows the flow of data between the spacecraft, the ground stations and the Roman Ground System.  GITL operations do require active ground contact to be present during the downlink and uplink portions of the activity, and the loop will idle until connection is re-established for either of these cases.

\begin{table}[ht]
\caption{Data downlinked during HOWFSC GITL operations} 
\label{tab:gitl_down}
\begin{center}       
\begin{tabular}{|p{0.4\linewidth}|p{0.1\linewidth}|p{0.4\linewidth}|}
\hline
Data product & Size & Quantity \\
\hline
HOWFSC GITL image, cleaned and cropped & $153\times153$ array & 21 or 35: 7 for each wavelength (3 probed pairs and one unprobed); imaging configurations use 3 wavelengths, while spectroscopy uses 5 \\
\hline
DM1 setting, absolute & $48 \times 48$ array & 21 or 35 \\
\hline
DM2 setting, absolute & $48 \times 48$ array & 21 or 35 \\
\hline
Ancillary telemetry: EXCAM gain & scalar	& 21 or 35 \\
\hline
Ancillary telemetry: EXCAM exposure time & scalar & 21 or 35 \\
\hline
Ancillary telemetry: number of EXCAM frames & scalar	& 21 or 35 \\
\hline
Ancillary telemetry: Color Filter Alignment Mechanism (CFAM) position (maps to current color filter) & 2 scalars & 21 or 35 \\
\hline
Ancillary telemetry: GITL crop region & 4 scalars & 21 or 35 \\
\hline
\end{tabular}
\end{center}
\end{table} 

\begin{table}[ht]
\caption{Data uplinked during HOWFSC GITL operations} 
\label{tab:gitl_up}
\begin{center}       
\begin{tabular}{|p{0.4\linewidth}|p{0.1\linewidth}|p{0.4\linewidth}|}
\hline
Data product & Size & Quantity \\
\hline
New DM1 setting, absolute & $48 \times 48$ array & 1 \\
\hline
New DM2 setting, absolute & $48 \times 48$ array & 1 \\
\hline
Probe scale factor FSW parameter & scalar & 6 (3 probes, positive and negative) \\
\hline
EXCAM gain FSW parameter & scalar & 12 or 20: 4 for each wavelength (1 for unprobed and 1 for each probe pair, both positive and negative); imaging configurations use 3 wavelengths, while spectroscopy uses 5 \\
\hline
EXCAM exposure time FSW parameter & scalar & 12 or 20 \\
\hline
EXCAM number-of-frames FSW parameter & scalar & 12 or 20 \\
\hline
Status FSW parameter (telemetered to tell the plan for the observation how to proceed) & scalar & 1 \\
\hline
Start FSW parameter (telemetered to actually make the plan for the observation continue) & scalar & 1 \\
\hline
\end{tabular}
\end{center}
\end{table} 

The GITL implementation choice was due to very practical considerations: HOWFSC was originally baselined to be implemented entirely in CGI flight software (FSW).  To address the extensive computational needs of HOWFSC, particularly for Jacobian computation, within the limits of space-qualified hardware, CGI had added an entire separate processor board (CPU + attached FPGA and memory) whose only purpose was to support HOWFSC and related calculations.  It also baselined a dedicated solid-state-recorder to store Jacobians in non-volatile memory, as they could be multiple gigabytes in size (depending on optical configuration) and take many tens of hours to calculate with the on-board processor, even with FPGA acceleration.  The timing needs were handled operationally, by using only a single Jacobian and running the calculation of it in the background while the WFI instrument is collecting data.

However, at the Roman Mission Preliminary Design Review (PDR) in late 2019, the timeline to develop the flight software for HOWFSC was flagged as a red risk, and a subsequent tiger-team investigation recommended the switch to a GITL approach.  In the end, this provided several architectural benefits for CGI:
\begin{itemize}
\item Risk: It did indeed reduce the FSW development risk, particularly as CGI took the opportunity to spin up a ground-software (GSW) element and offloaded every piece of functionality that was not essential to have on board into GSW.
\item Resources: The separate processor board and solid-state-recorder were subsequently descoped, saving mass, power, FSW development effort and overall system complexity.
\item Software development: The HOWFSC implementation itself was simplified as it could be implemented by WFSC subject matter experts in a modern high-level language   (Python in this case) with a robust ecosystem for numerical processing.  (The set of personnel who are knowledgeable about HOWFSC algorithms and the set of people who are suitably-trained to write flight software are largely disjoint, which creates staffing challenges.)
\item Computation speed: Offloading the computationally-expensive parts to the ground permitted that computation to be done on modern COTS hardware, which is significantly more capable than space-qualified hardware and not constrained by mass or power.  The requirement for the HOWFSC computation only, excluding data transfer, is 60 seconds, which during CGI TVAC was met handily.  Jacobian storage issues also become largely moot as off-the-shelf storage is inexpensive.
\item Flexibility: late in development, the realities of the built DM hardware--including low-order surface deformation, nonfunctional actuators, and actuator crosstalk--led to a change in HOWFSC approach where the Jacobian was now to be recomputed every iteration.  (Previous modeling \mycitep{Zho19} had indicated that this recalculation was not needed to meet requirements, and so it was not baselined.)  With GITL, we adopted the ``early Jacobian'' approach discussed in Sec. \ref{subsec:jac}, where we kicked off a new Jacobian for the next iteration as soon as the DM settings were available from the current one.  Had HOWFSC remained in FSW, an update like this would have been impossible, and we would have been forced to accept the performance degradation.
\end{itemize}
Looking forward to HWO, advances in algorithmic approaches since 2019 may have partially-obviated the need for some of the hardware choices CGI planned  (e.g. \mycitet{Wil21}) had they been available earlier, particularly the non-volatile storage.  However, computational cost for HOWFSC still grows as a high power of DM size:--the computation time for different parts of the HOWFSC pipeline grows as $N_{act}^4-N_{act}^6$ for an $N_{act} \times N_{act}$ DM\mycitep{Pog22}--and radiation-hardened Class A processors lag the COTS state-of-the-art by 10 years or more \mycitep{Bel23}.  Careful thought should be given to how much functionality needs to remain on-board vs off-spacecraft in future applications.

\subsection{Implementation of GITL} \label{subsec:gitl_imp}

\noindent In the transition to GITL, nearly all of the computation associated with HOWFSC was offloaded to the ground: all wavefront estimation, all wavefront control, all Jacobian calculation, and all steps to prepare CGI settings for the next iteration.  The exception was the data processing pipeline to perform basic image correction on frames before use: bias subtraction, cosmic-ray masking, dark subtraction, and correction for EM gain and flat fielding.  This functionality had to remain within the spacecraft as several other FSW capabilities also rely on EXCAM ``cleaned frames'' and also needed it present: fine alignment of focal-plane masks and field stops based on the speckle-balance technique \mycitep{Rig23} and automated acquisition of stars with EXCAM.  Note: the onboard pipeline is ``thin'' relative to the ground data pipelines that will process science data.  It contains no corrections of detector nonlinearity, or traps, or compensation for ``smearing'' during readout (as EXCAM does not have a shutter), and the routines for bias subtraction and cosmic-ray masking are of a simpler design compared ones used for the ground pipelines.

For HOWFSC, frame cleaning was augmented with functionality that would mean-combine or median-combine several frames taken with the same optical configuration and crop the resulting combined frame, and merge it with a per-frame bad pixel map.  Individual frames are passed down with bad pixels in the frame indicated by NaNs.  This combine-and-crop step is absolutely necessary to reduce the downlinked data volume, as HOWFSC GITL has to be run through the Roman S-band housekeeping link in order to support real-time operations.  The S-band link is intended for receiving spacecraft and instrument telemetry and sending commands rather than transferring imaging data, and has a much smaller capacity (hundreds of kbps) than the Ka-band science channel (hundreds of Mbps) used for downlink of complete science products in a non-real-time manner.  The cropped images were chosen to be 153x153, down from 1024x1024 on cleaned frames, large enough to fit the full wide-field-of-view (WFOV) dark hole at the reddest wavelength in Band 4 (10\% band centered at 825nm, longest-wavelength broadband filter on CGI) without dropping information, while reducing the required S-band downlink data volume by 44x.

In either the pre- or post-GITL scenario, data collection for HOWFSC is run via the Roman Science Observation Sequence Engine (SOSE). This sequence engine runs predefined procedures (``procs'') written in FSTOL, a flight-only variant of the System Test and Operations Language (STOL) developed by Goddard Space Flight Center (GSFC) for their ground-system operation.  Procs may execute instrument commands, call other procs, or run other STOL functionality such as comparing telemetry or reading data from file.   CGI has a dedicated command in instrument FSW to collect HOWFSC data, combine and crop it, and send it to the ground via S-band.  Other HOWFSC operations (moving DMs for probing, changing color filters to capture chromatic variation, changing EXCAM settings) are handled with more generic CGI utility commands run in an appropriate order by SOSE in a HOWFSC proc.  These procs were used through CGI TVAC to collect data for HOWFSC in a flight-like manner.

On the ground, tools written by the Science Support Center (SSC) will re-assemble the raw packets into data arrays and scalars, and feed them into the HOWFSC GITL algorithm.  The HOWFSC GITL software is required to take no more than 60 seconds to complete, and this tool will:
\begin{itemize}
\item Convert the incoming image arrays into normalized intensity by dividing arrays of photoelectrons by the exposure time in seconds, and then scaling each by a model-derived peak flux in photoelectrons/sec.
\item Flag every pixel not mapped to a pixel in the model dark hole as bad, so we avoid wasting time estimating fields that will not get used.
\item Estimate the current contrast from measured unprobed images, for use with the control strategy.
\item For each probe pair and each wavelength, compute the phase associated with each probe pair using the HOWFSC optical model (see Sec. \ref{subsec:pp} for phase estimation and Sec. \ref{subsec:model} for the optical model).
\item For each wavelength, take the list of normalized-intensity frames ($2N_{pair}+1$ assuming $N_{pair}$ pairs) and the list of probe phases just computed, and calculate a complex-valued electric-field estimate at each pixel (see Sec. \ref{subsec:flag} for details on flagging bad wavefront estimates and Sec. \ref{subsec:est} for the estimation itself).  Store estimation products for offline analysis.
\item For each wavelength, compute the electric field for each pixel based on the HOWFSC optical model for the current iteration.
\item Using the current iteration number and the estimated current contrast, select relevant control strategy parameters and build per-pixel weighting matrices (see Sec. \ref{subsec:cs} for details of CGI control strategy implementation and Sec. \ref{subsec:ppw} for the implementation of per-pixel weights).
\item Using the current DM setting, build the per-actuator weighting matrix (Sec. \ref{subsec:paw}).
\item Using the per-pixel weighting matrix and the precomputed Jacobian, compute $J^T W_{E}^T W_{E} J^T$ (see Sec. \ref{subsec:jac}).
\item Using the measured electric-field, precomputed Jacobian, $J^T W_{E}^T W_{E} J^T$, the per-pixel and per-actuator weighting matrices, and the DM multiplicative gain from the control strategy, compute the next DM settings with EFC.  (see Sec. \ref{subsec:efc}).
\item Run additional checks on DM settings to enforce neighbor rules and zero out subnormal values.
\item For each wavelength, compute the electric field for each pixel based on the HOWFSC optical model for the upcoming iteration.
\item Compute the expected contrast for the next iteration by calculating \linebreak
$I_{measured} + \left(I_{model,next} - I_{model,current}\right)$.
\item Using the control strategy, select the probe height for the next iteration and populate outgoing FSW parameter values.
\item For each wavelength, using the expected contrast profile in the dark hole for the next iteration and the control strategy, compute probed and unprobed camera settings (gain, exposure time, number of frames) to populate outgoing FSW parameter values.
\item Using the set of camera settings and parametrically-captured overheads, compute the length of time required for the next iteration.
\item Return DM settings, FSW parameter values, and estimation data products.
\end{itemize}

\begin{figure}
\centering
\includegraphics[width=1\textwidth]{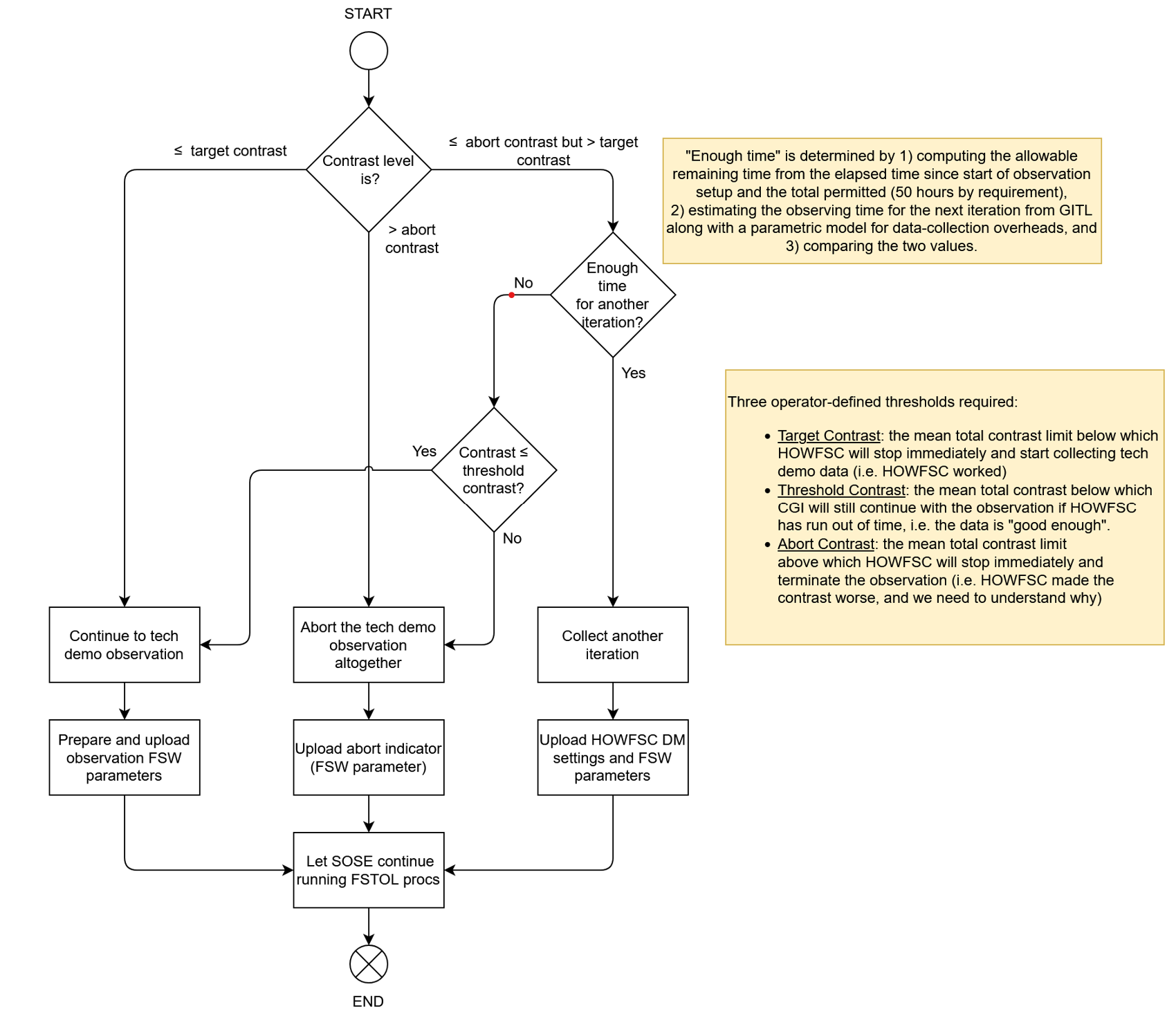}
\caption{Decision logic flow for iteration selection for HOWFSC, implemented by the Science Support Center (SSC) at Caltech/IPAC.} \label{fig:gitl_flow}
\end{figure}

Based on the expected contrast at the next iteration, and the time to complete, SSC will decide whether to continue to the observation, collect another iteration, or abort the observation entirely.  Figure \ref{fig:gitl_flow} shows the decision flow diagram, which can be automated given predefined decision thresholds.  Once uploads are complete, subsequent flow control is handled by FSTOL procs in SOSE and telemetered values of those FSW parameters set during the uplink.  The expected time for the entire loop is less than 30 minutes assuming ground contacts are available; if contact is unavailable, there may be up to a 4-5 hour gap until connection to the next ground station is established.  The contrast drift during a 44-hour observing period from beamwalk and structural and thermal modes of the telescope and instrument is estimated by modeling to be $< 1 \times 10^{-9}$ (see \mycitet{Kri23} for details), and so the lag from standard GITL or even a gap between stations is not expected to contribute significantly to the evolution of quasistatic speckles.

\section{Performance Results from TVAC}

\noindent Measuring contrast empirically is a critical part of verification of CGI performance requirements.  The resulting DM settings will also eventually form part of the initial seed for in-orbit commissioning.  Logistically, this has some challenges as CGI is designed fundamentally as a space instrument, to function in a vacuum with cold cameras.  (The 2014-2017 milestone tests were done in a vacuum chamber.)

Operation in atmosphere at ambient temperatures is possible, but ground-based telescopes employ dedicated fast adaptive optics systems to compensate for atmospheric fluctuations, and none of this functionality is available for CGI.  To make matters more challenging, the CGI DMs are very hygroscopic and change their shape significantly when they absorb moisture, which requires them to be placed in a continuous dry-nitrogen purge while in atmosphere--piping additional atmospheric turbulence into the middle of the instrument itself.  Instead, the approach we took was to do a Full Functional Test at ambient to shake out functional issues at the system level without performance expectations, and then run performance tests during TVAC testing, when the instrument was required to be in a vacuum chamber anyway.

This section describes the TVAC instrument integration and test campaign and its results.  All performance requirements introduced in Sec. \ref{sec:intro} were significantly exceeded.

\subsection{TVAC in the Integration and Test Context}

\noindent The final step in the implementation of CGI, or any other instrument at JPL, is the instrument integration and test (II\&T) phase.  This phase began in July 2022 with integration; during this period, the individual components of the instrument (cameras, DMs, electronics, static optics, thermal hardware, etc.) were delivered by the subsystems after verifying their lower-level requirements.  They were built up into larger assemblies (optical bench, thermal pallet with avionics, etc.) including optical alignment and wiring (``harness''), which in turn were assembled into the final instrument, an activity that finally completed in October 2023.

Following integration, the test portion of II\&T began, focused on closing system-level requirements on the instrument.  October and November of 2023 were spent in a Full Functional Test, which exercised all instrument functionality in flight and ground software which was possible to exercise in ambient conditions.  The majority of CGI flight software requirements were verified during this activity.  This test period also included running all calibration data collection and processing activities leading to the creation of a HOWFSC model, and the exercise of HOWFSC functionality, to ensure that the mechanics functioned and to minimize wasted time during TVAC due to bugs that could have been caught earlier.  No performance expectations were levied here, and unsurprisingly the calibrations were poor due to turbulence effects and HOWFSC did not null appreciably.

Following this, CGI entered environmental testing, which subjected the instrument to extreme environmental conditions to verify survival requirements in the presence of specified disturbances, such as vibration profiles or in the case of TVAC, flight-like hot and cold vacuum environmental conditions.  TVAC is also when the thermal system is tuned, and for CGI was also an opportunity to engage in performance testing while under vacuum.  The TVAC campaign for CGI was 54 days, 3 partially-overlapping 9-hour shifts a day, 7 days a week, and scheduled tightly due to the cost of running the chamber and paying for the standing army of supporting personnel.

II\&T TVAC was the first opportunity to do coronagraphy on the instrument in flight-like conditions, but it was also the last prior to launch.  The delivery date of CGI to Goddard Space Flight Center (GSFC) was a month after TVAC completion, just enough time to measure the instrument mass properties and complete paperwork.  For TVAC CGI had built the Coronagraph Verification Stimulus (CVS \mycitep{Bai23}), a dedicated piece of optical ground-support equipment (OGSE) which simulates both a star at flight-like magnitudes (generally $V_{mag}$ between 0 and 3 for HOWFSC) and the optical front end of the telescope.  No similar hardware exists in higher levels of I\&T, and CGI will not be able to see another star, simulated or otherwise, until in-orbit commissioning.  For this reason, a significant amount of effort was dedicated to HOWFSC and prerequisites such as calibrations to populate the HOWFSC optical model: 34 shifts of HOWFSC and 18 shifts of calibration spread across the 54-day period.

\subsection{TVAC Result Summary}

\begin{figure}
\centering
\includegraphics[width=1\textwidth]{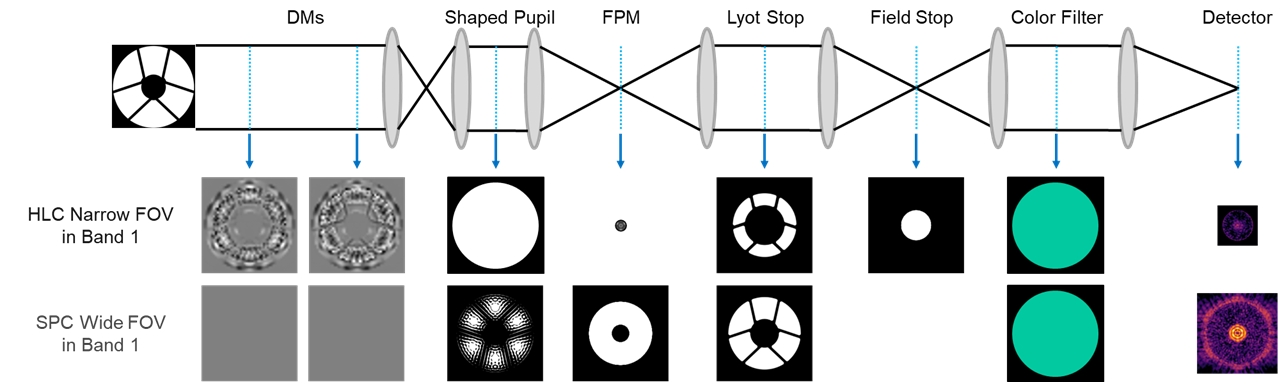}
\caption{Unfolded layouts of the NFOV HLC and WFOV SPC architectures, showing masks, DM settings, and filters.} \label{fig:layouts}
\end{figure}

\noindent By the end of TVAC, CGI had performed HOWFSC with two different mask configurations, both capable of meeting TTR5:
\begin{itemize}
\item A Hybrid Lyot Coronagraph (HLC\mycitep{Tra16}) uses a complex-valued focal-plane mask along with a Lyot stop and large-amplitude, structured patterns on both DMs to create a deep null that pushes all of the light just outside a desired radial extent.  In practice, a field stop is also required to handle the large dynamic range on the camera and prevent significant ghosting within the imaging lens.  This coronagraph does not use a pupil mask upstream of the focal-plane mask.  This HLC mask was optimized for narrow field-of-view (NFOV, $3-9 \lambda/D$) operation with coronagraph Band 1 (a 58nm band centered at 575nm).
\item A Shaped Pupil Coronagraph (SPC\mycitep{Zim16a}) uses a binary pupil-plane mask upstream of the focal plane, along with a binary focal plane mask that bounds the dark-hole extent and a Lyot stop.  As the bulk of the light suppression is done at the focal-plane mask and Lyot stop, SPC masks do not in general require a field stop to operate.  This SPC mask was optimized for wide field-of-view (WFOV, $6-20 \lambda/D$) operation with coronagraph Band 1.
\end{itemize}
The layouts of these two architectures are given Fig. \ref{fig:layouts}.  Further discussion of the details of the CGI mask designs can be found in \mycitet{Rig21}.  For both of these scenarios, HOWFSC was initialized with a model-based DM seed created by running the HOWFSC optical model against itself.

Table \ref{tab:tvac} gives the coherent and incoherent static raw contrast values achieved with both configurations.  In both cases, not only were both requirement levels achieved, but the total raw contrast for each was less than the smaller of the two requirements ($\leq5 \times 10^{-8}$), eliminating any possible concerns about distinguishing coherent signal from incoherent.  Note that raw contrast values:
\begin{enumerate}
\item include a spatially-varying correction factor applied to account for the reduced planet flux near the inner working angle, as compared to raw normalized intensity.
\item do not employ any postprocessing or PSF subtraction other than removal of detector noise as described in Sec. \ref{subsec:gitl_imp}, although they do average over a region.  Additional gains of $2\times$ or more may be expected from postprocessing, depending on PSF-subtraction method and detector noise properties \mycitep{Ygo21}
\end{enumerate}

\begin{table}[ht]
\caption{Performance results from TVAC performance testing} 
\label{tab:tvac}
\begin{center}       
\begin{tabular}{|p{0.36\linewidth}|p{0.18\linewidth}|p{0.18\linewidth}|p{0.18\linewidth}|}
\hline
Configuration & Requirement & NFOV HLC & WFOV SPC \\
\hline
Static Coherent Raw Contrast, $6-9 \lambda/D$, Band 1 & $\leq 5\times10^{-8}$ & $0.98 \times 10^{-8}$ & $1.42 \times 10^{-8}$ \\
\hline
Static Incoherent Raw Contrast, $6-9 \lambda/D$, Band 1 & $\leq 10^{-7}$ & $2.35 \times 10^{-8}$ & $2.96 \times 10^{-8}$ \\
\hline
Static Total Raw Contrast, $6-9 \lambda/D$, Band 1 & - & $3.33 \times 10^{-8}$ & $4.38 \times 10^{-8}$ \\
\hline
\end{tabular}
\end{center}
\end{table} 

\subsubsection{TVAC results: NFOV} \label{subsec:nfov}

\noindent The Band 1 NFOV configuration is the primary supported mode for CGI, and the expected configuration to be used for verifying performance consistent with the TTR5 requirement.  (See Table \ref{tab:nfov} for details.)  As such, it received a considerable amount of allocated TVAC time: 17 shifts of alignment and calibration activities as prerequisites for populating the HOWFSC model and processing the EXCAM data to remove detector noise, and 15 shifts solely for running HOWFSC.  After 15 shifts, the team reached a coherent contrast of $2.42 \times 10^{-8}$ and an incoherent contrast of $4.45 \times 10^{-8}$, sufficient to verify TTR5 on its own, and authorization from NASA HQ was granted to spend an additional 5 days (15 shifts) on HOWFSC to push limits of CGI performance further, for a total of 30 shifts.

\begin{table}[ht]
\caption{NFOV HOWFSC TVAC properties}
\label{tab:nfov}
\begin{center}       
\begin{tabular}{|p{0.4\linewidth}|p{0.4\linewidth}|}
\hline
Property & Value \\
\hline
Band of operation & Band 1 (10\% band centered at 575nm) \\
Number of subbands & 3 \\
Subfilters used & Bands 1A, 1B, 1C \\
Equivalent V-band magnitude of CVS & V = 3.05 \\
Number of wavelengths used in model & 3 (tied to effective center wavelengths of subfilters) \\
Inner radius of dark hole & 2.9 $\lambda/D$ at $\lambda = 546$nm (blue edge of Band 1) \\
Outer radius of dark hole & 9.1 $\lambda/D$ at $\lambda = 604$nm (red edge of Band 1) \\
\hline
\end{tabular}
\end{center}
\end{table} 

This additional time proved very fruitful, as it provided the opportunity to diagnose an unexpected behavior that had appeared: wavefront tilt was building up in the system during iterations, driving the LOWFSC tilt setpoint eventually as high as 23nm off-center.  Analysis determined that this was due to regions of low phase near the edge of the pupil in the phase retrieval that was used as the upstream wavefront in the control model.  These regions were biasing the tip-tilt removal, and causing a shift in the PSF in the model.  The calculations for the model-based DM seed attempted to recenter the PSF by adding a tilt to the initial DM settings, and iterating on the instrument exacerbated the problem.  This model mismatch also slowed down the convergence rate of HOWFSC.

The implemented fix was to 1) adjust the tip-tilt removal in the upstream phase retrieval so that the Fourier transform gives a centered PSF, and update the HOWFSC model, 2) create a new model-based DM seed which removes any residual tip/tilt before applying it to the instrument, and 3) restart HOWFSC nulling from scratch at the new model seed, in a dedicated second run.

\begin{figure}
\centering
\includegraphics[width=0.75\textwidth]{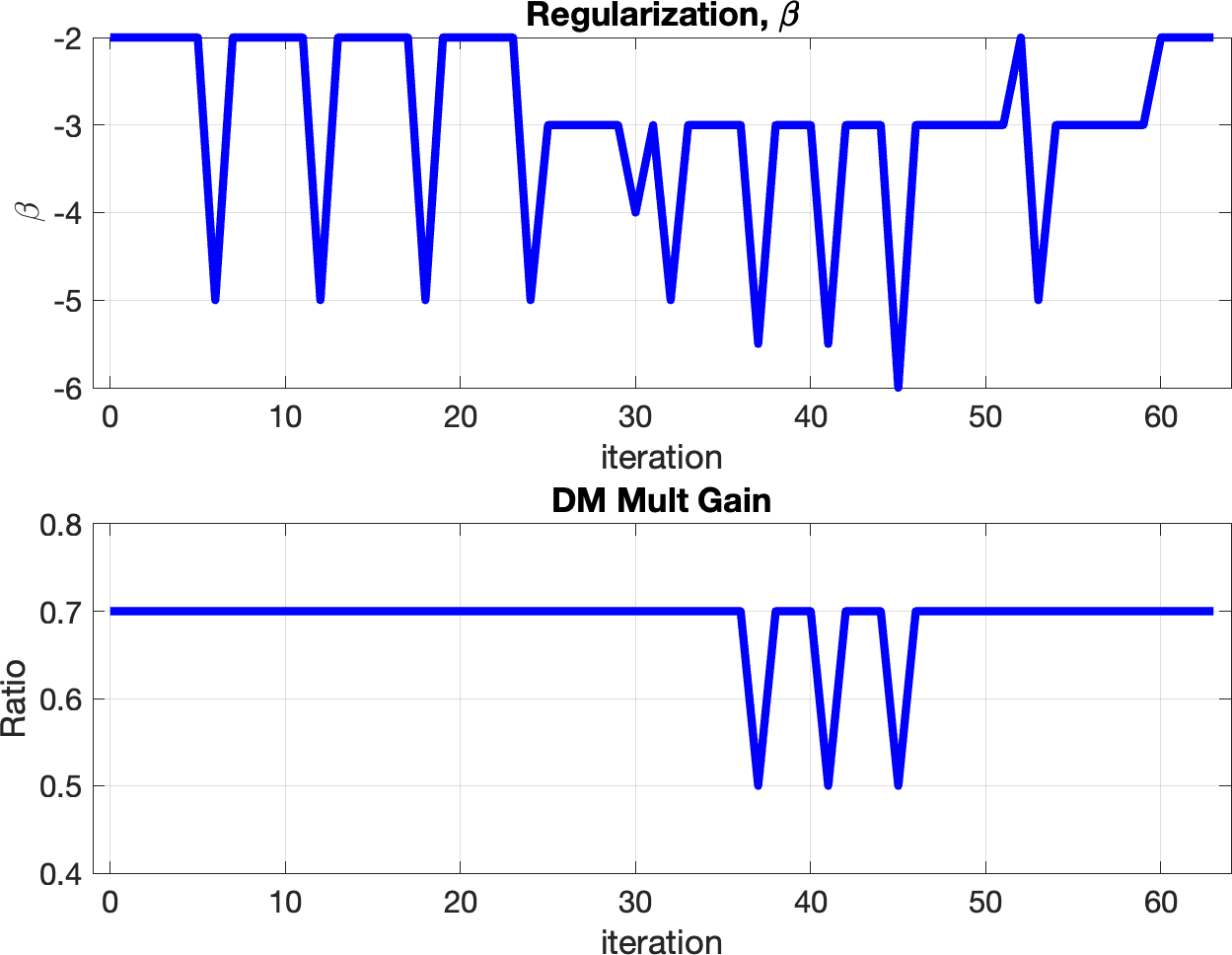}
\caption{Control strategy variables (regularization parameter $\beta$, multiplicative DM gain) as a function of iteration during Band 1 NFOV Run \#2.} \label{fig:bd_nfov}
\end{figure}

Trying again from scratch turned out to be successful; without the model mismatch, NFOV converged immediately and very smoothly.  Figure/Video \ref{vid:nfov} shows the contrast vs. iteration for the second NFOV run, beginning from scratch.  Figure \ref{fig:bd_nfov} shows the control strategy variables used; we adopted an aggressive beta-bumping schedule near the end, followed by a period with no beta-bumps near the end of the allotted time to ensure we had settled completely before evaluating performance.  In addition to exposure time, each Band 1 iteration takes approximately 22 minutes for overheads.

\begin{video}
\begin{center}
{\includegraphics[width=1\textwidth]{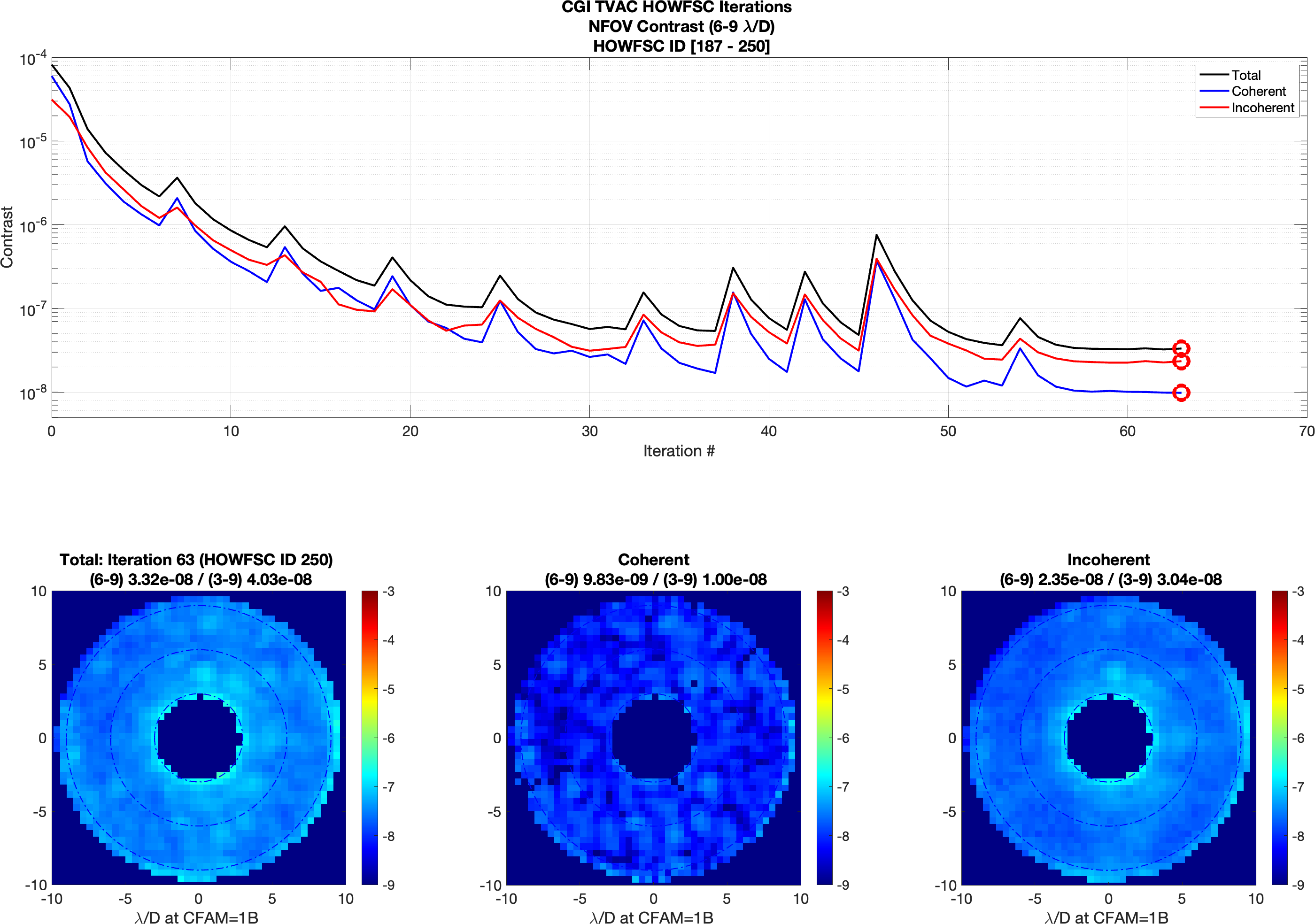}}
\\
\end{center}
\caption{\label{vid:nfov}This image is the final frame from Video 1 (Video 1, AVI, 31.4 MB).  \textit{Top.} Average contrast in the $6-9 \lambda/D$ annulus as a function of iteration.  \textit{Bottom.} Total (\textit{left}), Coherent (\textit{center}), and Incoherent (\textit{right}) contrast in the Band 1 NFOV configuration at the iteration marked on the top curve.}
\end{video}

\begin{figure}
\centering
\includegraphics[width=0.5\textwidth]{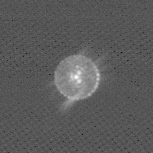}
\caption{EXCAM image of the dark hole in band 1B (a 3.3\%-band engineering filter centered at 575nm) during HOWFSC iteration ID 250 (final iteration of NFOV run \#2), showing incoherent ``tails'' ultimately determined to be due to a light leak at CFAM.  Logarithmic scale used to improve tail visibility.} \label{fig:it250}
\end{figure}

One issue that became apparent near the end of run \#2 was an unexpected amount of incoherent light setting a floor in NFOV observations.  Figure \ref{fig:it250} shows an image of the unprobed dark hole at the final iteration of run \#2, and some ``tails'' are clearly visible extending beyond the edge of the field stop.  Subsequent investigation revealed that the NFOV field stop, when combined with the rest of the coronagraph to make a dark hole, was introducing high-angle diffracted light in the near-pupil collimated space where the Color Filter Alignment Mechanism (CFAM) is placed, and some of that light was making it around the edge of the filter.  This mechanism was unfortunately not caught in stray light testing earlier as it does not appear at appreciable levels when the DMs + coronagraph are not in a dark hole configuration.

This particular location, fortunately, had been designed to mount a baffle even though none was installed, and both the mounting points in the table and baffle mount itself were already extant.  A baffle was fabricated and mounted after the completion of TVAC, and inspection with an alignment scope indicated that the sneak path was successfully blocked.  We do not expect this specific incoherent signal to appear again in flight.

The NFOV incoherent floor was partially set by this stray light; aside from this, however, the coherent signal continued to drop with aggressive regularization.  We expect that additional iterations would have yielded a lower coherent floor, and we are confident that the installation of the baffle will result in some reduction of the incoherent floor during flight.  However, we also expect from modeling that the incoherent tails do not constitute the entirety of the incoherent signal; see discussion in Sec. \ref{subsec:ll} for future directions.

NFOV was also the venue for a ``scenario test'', a validation exercise to run the coronagraph in the manner intended during observation, to ensure we are doing the correct activities in the correct order.  This activity began directly after the completion of Run \#2, and started by pulling all coronagraph masks out of the beam, misaligning the star, and injecting a realistic line-of-sight jitter profile through the CVS at the 6-7mas rms level.  Then, using only procs intended for flight use, we performed EXCAM acquisition, mask insertion and alignment, LOWFSC self-calibration, and executed HOWFSC.  HOWFSC recovered the performance from the end of run \#2 within 3 iterations; closed-loop jitter residuals were $<0.5$ mas rms.  The test subsequently exercised a simulated short CGI observation sequence successfully.

\subsubsection{TVAC results: WFOV} \label{subsec:wfov}

\noindent WFOV testing was done as a ``target of opportunity'', taking advantage of additional available downtime in the TVAC schedule during thermal testing to exercise an additional coronagraph configuration. Band 1 WFOV, previously an unsupported mode contributed by the NASA Exoplanet Exploration Program, was selected for 3 reasons:
\begin{itemize}
\item First and most important, Band 1 WFOV is the only other configuration that ships with CGI that is capable of demonstrating TTR5, and demonstrating capability with two independent architectures reduces the risk that CGI is unable to provide functionality at TTR5 levels during flight operations.
\item More practically, doing a second observation in Band 1 allowed Band 1 NFOV calibration data to be largely reused, minimizing the work to commission a second mode.  Properties are given in Table \ref{tab:wfov}.
\item There was a preference to test HOWFSC with at least one instance of a shaped-pupil coronagraph given the architectural differences seen in Fig. \ref{fig:layouts}.  While different architectures pose additional challenges in implementation, they also provide robustness in addition to extending capability: due to the lack of a field stop, CGI shaped pupil architectures are immune to the CFAM stray light issues discussed for NFOV, and could have satisfied TTR5 even if the baffle had not been installed.
\end{itemize}

\begin{table}[ht]
\caption{WFOV HOWFSC TVAC properties}
\label{tab:wfov}
\begin{center}       
\begin{tabular}{|p{0.4\linewidth}|p{0.4\linewidth}|}
\hline
Property & Value \\
\hline
Band of operation & Band 1 (10\% band centered at 575nm) \\
Number of subbands & 3 \\
Subfilters used & Bands 1A, 1B, 1C \\
Equivalent V-band magnitude of CVS & V = -1.710 \\
Number of wavelengths used in model & 3 (tied to effective center wavelengths of subfilters) \\
Inner radius of dark hole & 5.81 $\lambda/D$ at $\lambda = 546$nm (blue edge of Band 1) \\
Outer radius of dark hole & 19.15 $\lambda/D$ at $\lambda = 604$nm (red edge of Band 1) \\
\hline
\end{tabular}
\end{center}
\end{table} 

\begin{figure}
\centering
\includegraphics[width=0.75\textwidth]{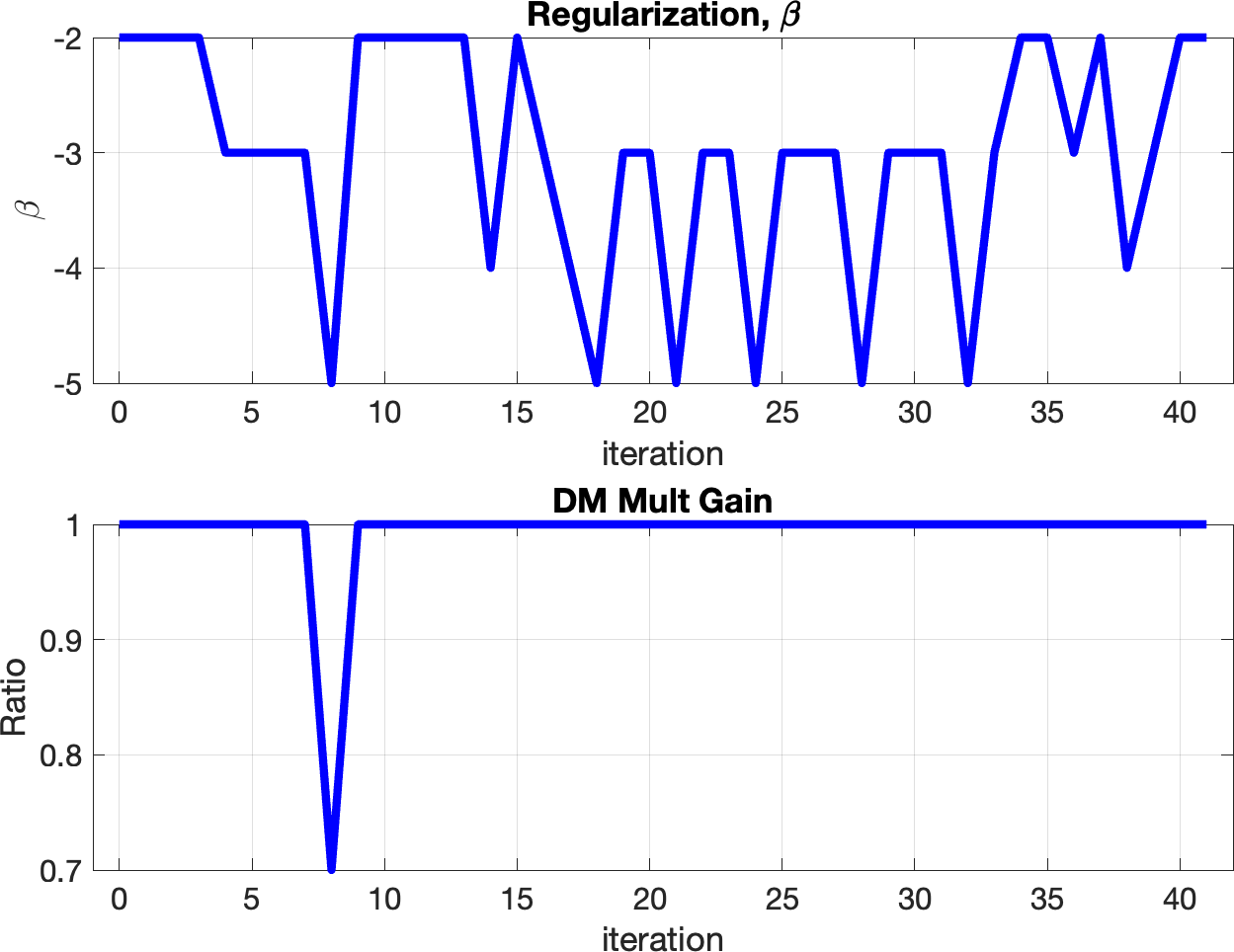}
\caption{Control strategy variables (regularization parameter $\beta$, multiplicative DM gain) as a function of iteration during Band 1 WFOV.} \label{fig:bd_wfov}
\end{figure}

Figure/Video \ref{vid:wfov} shows the contrast vs. iteration for WFOV operations, and Fig. \ref{fig:bd_wfov} shows the evolution of control strategy variables during iteration.  WFOV used 1 shift for dedicated calibration (along with reuse of the common-mode calibration work done for Band 1 NFOV, such as wavefront flattening and measurement of detector-to-mechanism transformations) and 4 total non-contiguous shifts of nulling operations (split into 1 and 3).  The gap occurs between iterations 11 and 12 in Figure/Video \ref{vid:wfov}, with several days of thermal characterization--and no mechanism motions--in between.  We note in particular that WFOV nulling ended entirely due to lack of available time; the contrast remains on a downward trend, and there does not seem to be an indication that the achieved performance represents any sort of floor in CGI capability.

\begin{video}
\begin{center}
{\includegraphics[width=1\textwidth]{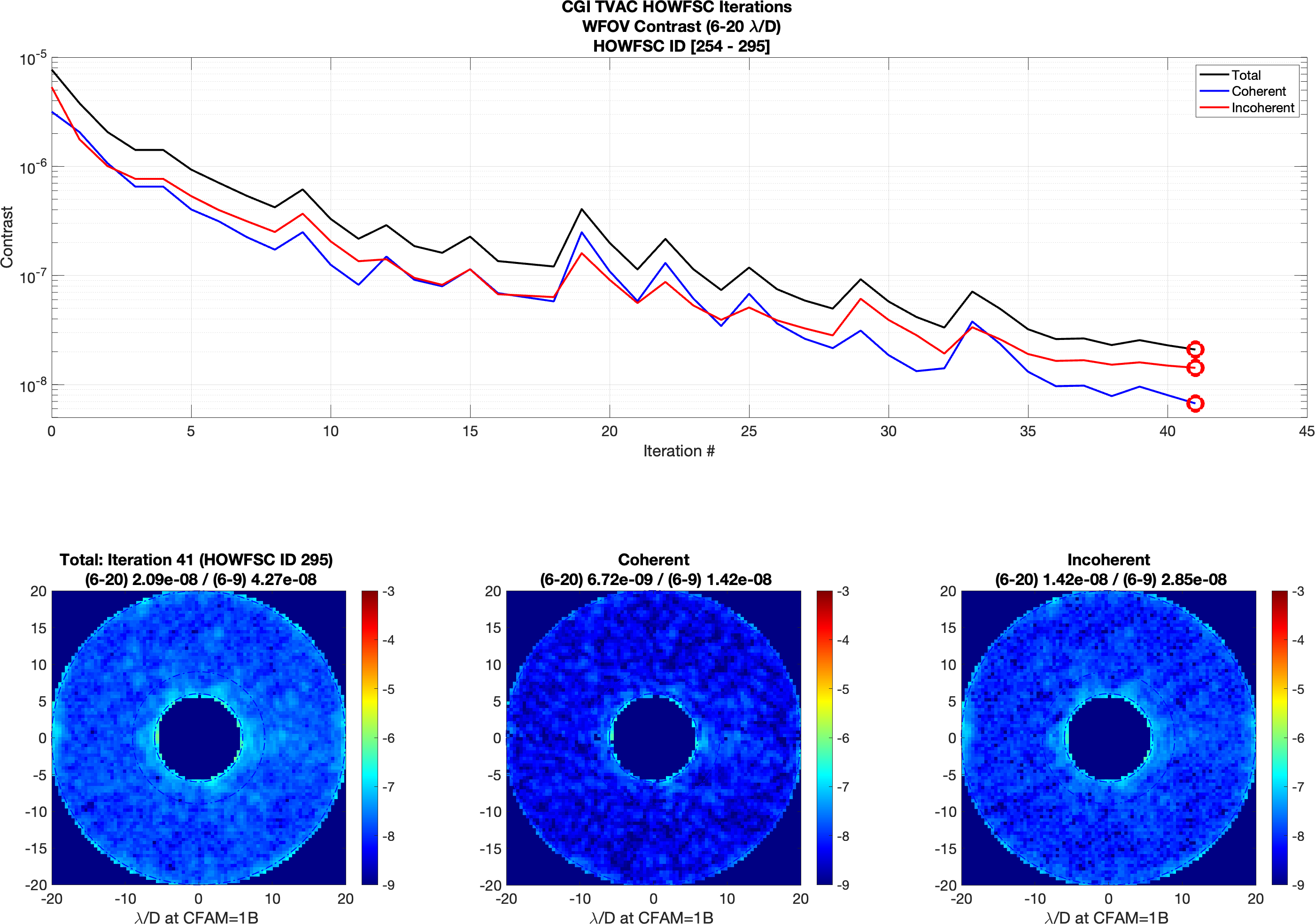}}
\\
\end{center}
\caption{\label{vid:wfov}This image is the final frame from Video 2 (Video 2, AVI, 20.3 MB).  \textit{Top.} Average contrast in the $6-9 \lambda/D$ annulus as a function of iteration.  \textit{Bottom.} Total (\textit{left}), Coherent (\textit{center}), and Incoherent (\textit{right}) contrast in the Band 1 WFOV configuration at the iteration marked on the top curve.}
\end{video}

\subsection{Examination of Beta Bumps with the SVD Spectrum}

\begin{figure}
\centering
\includegraphics[width=1\textwidth]{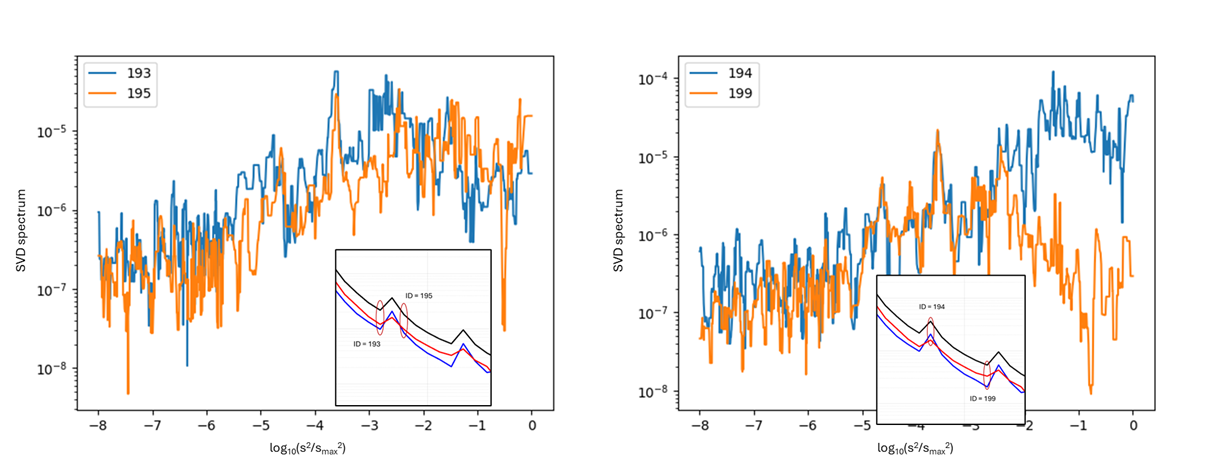}
\caption{\textit{Left.} Comparison of the SVD mode spectrum before and after a $\beta = -5$ bump.  \textit{Right.} Comparison of the SVD mode spectrum before and after 5 iterations of $\beta = -2$ control.  The horizontal axis is in the same units as $\beta$.  Note that the curves are run through a 9-element median filter for plotting legibility.} \label{fig:193-199}
\end{figure}

\noindent One useful tool for examination what HOWFSC iterations actually did is to look at the relative strength of various modes of the singular value decomposition with a singular-mode spectrum \mycitep{Sid17}.  Equation \ref{eq:USVT} noted that we can decompose the Jacobian via SVD into $U S V^T$.  If we compute $U^T W_{E} \mathbf{E}$, this gives the projection of the measured electric field into each SVD mode, which then map to $\beta$ values according to the elements of $S$.  For convenience, it is easier to plot $|U^T W_{E} \mathbf{E}|^2$ and remain real-valued.  Examples are shown in Fig. \ref{fig:193-199}.

The first thing to note is that ``same contrast'' does not imply ``same modal content''.   The left side of Fig. \ref{fig:193-199} shows two iterations directly before and after a beta bump with $\beta = -5$, with roughly the same contrast--but after the bump, there is considerably less power in modes corresponding to larger $\beta$ values: $3.6\times$ less power between -5 and -3.  Compare with the right side, which is entirely made of iterations with $\beta = -2$: the power in the ``easy modes'' are driven downward sharply ($17.8\times$ less power between -3 and 0), but the modes up to $\beta = -5$ are nearly untouched ($\approx 7\%$ change in power between -5 and -3).  This is the effect we expected from the discussion in Sec. \ref{subsec:bb}: only the periods of aggressive regularization actually touch the power at small $\beta$, and this is both demonstration on why we find beta-bumping necessary, and confirmation that it is working as intended.

\begin{figure}
\centering
\includegraphics[width=1\textwidth]{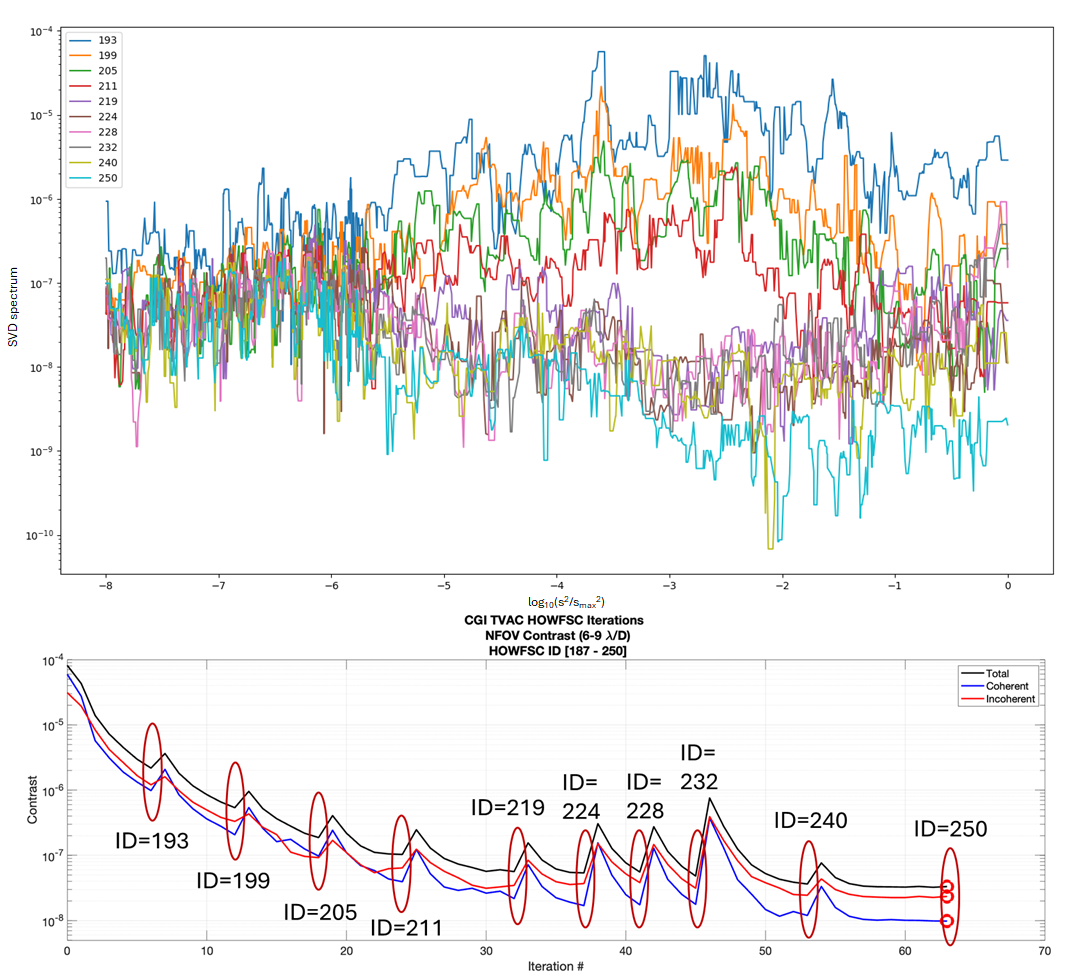}
\caption{\textit{Top.} SVD mode spectrum for 10 iterations collected during NFOV run \#2. The horizontal axis is in the same units as $\beta$.   Note that the curves are run through a 9-element median filter for plotting legibility.  \textit{Bottom.} Contrast plot from Figure/Video \ref{vid:nfov} showing the locations of the 10 iterations in the sequence.} \label{fig:10bumps}
\end{figure}

\begin{figure}
\centering
\includegraphics[width=1\textwidth]{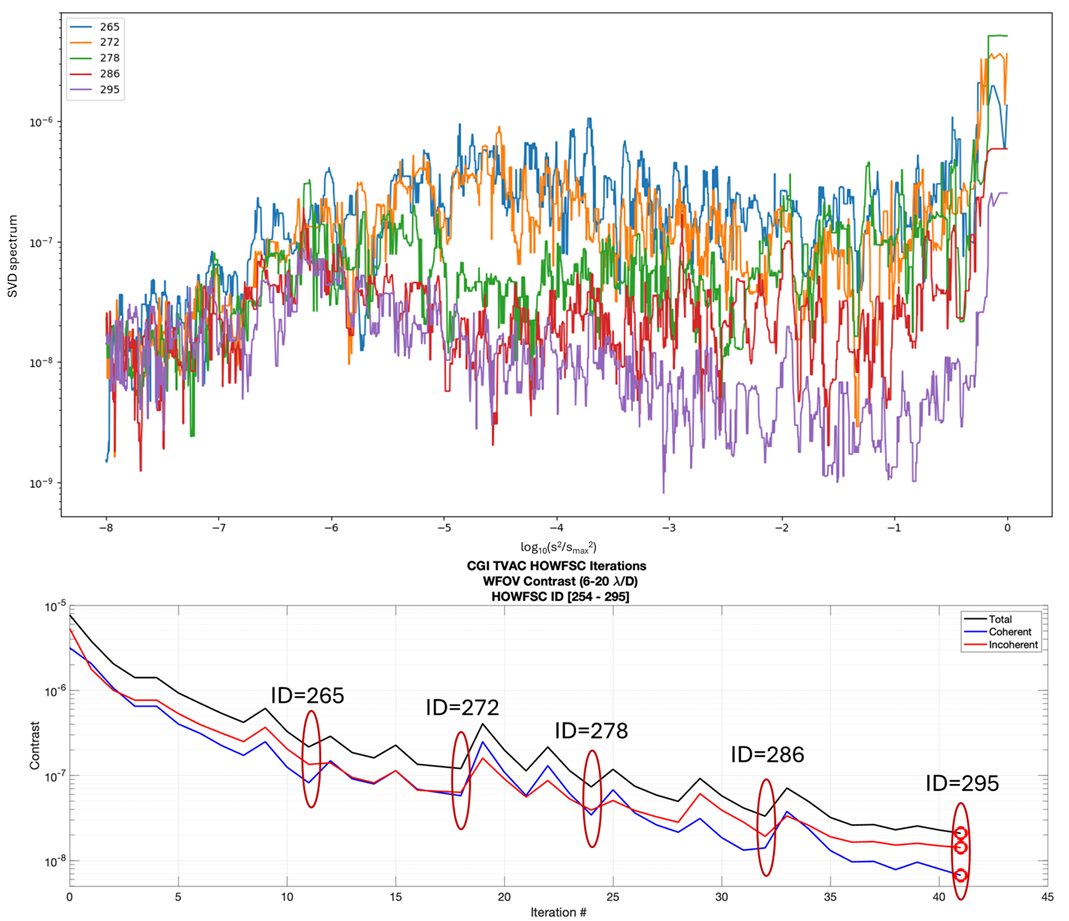}
\caption{\textit{Top.} SVD mode spectrum for 5 iterations collected during WFOV.  The horizontal axis is in the same units as $\beta$.  Note that the curves are run through a 17-element median filter for plotting legibility.  \textit{Bottom.} Contrast plot from Figure/Video \ref{vid:wfov} showing the locations of the 5 iterations in the sequence.} \label{fig:5bumps}
\end{figure}

Figure \ref{fig:10bumps} shows the SVD spectrum at 10 iterations through the NFOV run \#2 iteration sequence.  The benefit of multiple beta bumps to the overall contrast floor is clear.  This SVD analysis is also one reason why we are confident in asserting that these results do not represent a fundamental contrast floor yet--power remains in modes accessible by the range of tested $\beta$ values.  A similar plot for WFOV is shown in Fig. \ref{fig:5bumps}.

\subsection{Other Lessons Learned and Open Issues} \label{subsec:ll}

\noindent Beyond the concerns outlined above, such as the wavefront-tilt accumulation and the light leak around CFAM, several additional open issues and lessons learned during TVAC are worth commenting on:

\begin{itemize}
\item \textit{Convergence}: In the last several iterations of NFOV, the contrast convergence appears to stall, but that was an intentional decision. To ensure that we had fully cleaned up the dark hole after the last aggressive regularization, we conservatively let HOWFSC run with mild regularizations until we were sure contrast had fully converged to a local minimum for reporting purposes.  While this choice was effective for this specific goal, we likely left performance gains on the table by not squeezing in additional beta-bumps.  In the future, a control strategy better tuned to the length of nulling could provide additional gains.

\item \textit{Aggressive $\beta$ scheduling}: As discussed in \mycitet{Mar17}, selecting a $\beta$ which is too aggressive may result in no real improvement in an iteration due to calibration error and nonlinearities neglected in the linear EFC formulation.  But not selecting aggressive-enough betas will definitely \textit{never} allow the harder modes to be corrected.  It was appreciated late that we could have been bumping more aggressively.  We only went to $\beta = -6$ a single time, in NFOV run \#2, with a multiplicative gain on the correction of 0.5.  WFOV was never run with $\beta = -6$ at all.  We can see the impact in Figs. \ref{fig:10bumps} and \ref{fig:5bumps}, each of which has most of its remaining modal power below $\beta = -5$ by the end of the iteration.

\item \textit{Less-aggressive $\beta$ scheduling}: Both NFOV and WFOV used $\beta = -2$ and $-3$ for cleanup after $\beta$ bumps (Figs. \ref{fig:bd_nfov} and \ref{fig:bd_wfov}), and it shows in Figs. \ref{fig:10bumps} and \ref{fig:5bumps}.  But we did not schedule in cleanup iterations in the $\beta = -4$ vicinity, and this clearly shows as well at later iterations.  More attention to the full range of SVD modes would help ensure a lower floor.

\item \textit{Photometry}: Photometric calibration, which is necessary for conversion to normalized intensity and then to contrast, was a challenge during TVAC as the CVS light source exhibited photometric and spectral variability over time.  The baseline HOWFSC approach (estimate flux with a system throughput model and a model of the input magnitude and spectrum) was sufficient to get through HOWFSC, but we did apply correction factors sourced from LOCAM (Low Order CAMera = CGI wavefront sensor camera) measurements before reporting final contrast values.  LOCAM uses the same EMCCD detector type as EXCAM and standard LOCAM telemetry includes a record of the total number of counts received, and this could be used to monitor variability of the CVS over time.

This is not expected to have the same impact during operations, as measurement of reference-star photometry and calibration of absolute flux against standard stars are planned calibration activities for CGI\mycitep{Zel22}.

\item \textit{Probe swapping}: During the first NFOV run, we found that the control had regions of low wavefront modulation by some probes, such that some light was being erroneously flagged as incoherent.  Quick real-time fixes were made to swap out the DM probes for probes with different centers $(x_{c}, y_{c})$ or clocking angles $\theta$, and this change was sufficient to permit those focal-plane regions to be modulated and the associated speckles to be nulled.  This was not something that was seen during 2014-2017 Milestone testing, and so it remains a surprise.  However, the first NFOV run was also plagued by tilt offsets, and we have not yet determined how, if at all, that played into the appearance of these low-modulation regions.  We did continue to do this swapping during the second NFOV run (Sec. \ref{subsec:nfov}) and the WFOV run (Sec. \ref{subsec:wfov}).  

There remains open tasks to determine cause and what, if any action to take.  If it turns out to be an effect of the wavefront tilt, likely no action specific to this issue will be taken.  Otherwise, we may update the default probe shapes for each coronagraph configuration, or worst-case revisit the current concept of operations, which uses a single probe pattern for an entire HOWFSC run on a particular target.  These open tasks will be closed before launch, and are not expected to require on-orbit time to address.

\item \textit{Reduction of incoherent signal}: As mentioned in Sec. \ref{subsec:nfov}, some of the NFOV incoherent signal can be attributed to stray light leaking around CFAM, but not all, and no incoherent light in WFOV is attributable to this cause.  While the current contrast level does meet requirements, there is an open modeling and analysis task to analyze the TVAC data further and see other sources of incoherent can be identified and remedied if possible.

We also note that for most of the contrast curves in Figs. \ref{vid:nfov} and \ref{vid:wfov}, the incoherent curve follows the total contrast closely, a behavior inconsistent with the static incoherent formulation in Eq. \ref{eq:iinc}.  This is attributable to underestimation of the coherent intensity in Eq. \ref{eq:iinc2}; root cause for the underestimation has been a longstanding HOWFSC challenge even prior to CGI (e.g. \mycitet{Cad14}).  It is reasonable to surmise that improvements in estimating the coherent/incoherent split may improve nulling efficiency.  This uncertainty is also why TVAC tests pushed so hard to get total contrast below both the coherent and incoherent allocations.

\item \textit{Lack of convergence of WFOV ``easy'' modes}: the ``easiest'' modes in WFOV--the ones that should correct with the least applied stroke--start much higher than the rest of the curve in Fig. \ref{fig:5bumps} (see the far-right edge).  While the modes drop with iteration, they do not drop as far as the rest of the adjacent curve.  There is an open item to determine what is keeping these modes from being corrected as effectively as they should be.   Note that this behavior was not seen in NFOV.
\end{itemize}

\section{Summary}

\noindent The HOWFSC architecture for CGI has been defined and implemented in ground and flight software, and successfully executed the first time while undergoing performance testing in a flight-like environment with stiff time constraints in spring 2024 during the instrument TVAC campaign.  CGI exceeded the contrast required for its top-level requirement, TTR5, by $\geq 3\times$, with two independent coronagraph architectures covering $3-9$ and $6-20 \lambda/D$ with a $360^{\circ}$ dark hole on both.  The ultimate performance floor of either configuration is not known; contrast was limited by available time and partly by a since-repaired stray light leak for NFOV.  This robust, promising result makes a strong case that CGI will perform comparably on astrophysical targets in flight, providing a dataset with some of the highest contrasts available until the eventual implementation of HWO.

\section*{Disclosures}
\noindent The authors declare no conflicts of interest.

\section*{Code and Data Availability}
\noindent Release of codes used in this work is governed by JPL and NASA rules which limit dissemination scope and methodology.  The corresponding author should be contacted about the availability of any codes or data used in this paper.

\section*{Acknowledgments}

\noindent The research was carried out at the Jet Propulsion Laboratory, California Institute of Technology, under a contract with the National Aeronautics and Space Administration.   This work was authored by employees of Caltech/IPAC under Contract No. 80GSFC21R0032 with the National Aeronautics and Space Administration.

\bibliography{refs}

\begin{thebibliography}{10}

\bibitem{Men22}
B.~Mennesson, V.~P. Bailey, R.~Zellem, {\em et~al.}, ``{The Roman Space
  Telescope coronagraph technology demonstration: current status and relevance
  to future missions},'' in {\em Space Telescopes and Instrumentation 2022:
  Optical, Infrared, and Millimeter Wave},  L.~E. Coyle, S.~Matsuura, and M.~D.
  Perrin, Eds.,  {\bf 12180}, 121801W, International Society for Optics and
  Photonics, SPIE  (2022).

\bibitem{Pob22}
I.~Poberezhskiy, K.~Heydorff, T.~Luchik, {\em et~al.}, ``{Roman coronagraph
  instrument: engineering overview and status},'' in {\em Space Telescopes and
  Instrumentation 2022: Optical, Infrared, and Millimeter Wave},  L.~E. Coyle,
  S.~Matsuura, and M.~D. Perrin, Eds.,  {\bf 12180}, 121801X, International
  Society for Optics and Photonics, SPIE  (2022).

\bibitem{Fei24}
L.~Feinberg, J.~Ziemer, M.~Ansdell, {\em et~al.}, ``{The Habitable Worlds
  Observatory engineering view: status, plans, and opportunities},'' in {\em
  Space Telescopes and Instrumentation 2024: Optical, Infrared, and Millimeter
  Wave},  L.~E. Coyle, S.~Matsuura, and M.~D. Perrin, Eds.,  {\bf 13092},
  130921N, International Society for Optics and Photonics, SPIE  (2024).

\bibitem{Gal23}
R.~Galicher and J.~Mazoyer, ``Imaging exoplanets with coronagraphic
  instruments,'' {\em Comptes Rendus. Physique} {\bf 24}(S2), 69--113  (2023).

\bibitem{Nem23}
B.~Nemati, J.~Krist, I.~Poberezhskiy, {\em et~al.}, ``{Analytical performance
  model and error budget for the Roman coronagraph instrument},'' {\em Journal
  of Astronomical Telescopes, Instruments, and Systems} {\bf 9}(3), 034007
  (2023).

\bibitem{Kri23}
J.~E. Krist, J.~B. Steeves, B.~D. Dube, {\em et~al.}, ``{End-to-end numerical
  modeling of the Roman Space Telescope coronagraph},'' {\em Journal of
  Astronomical Telescopes, Instruments, and Systems} {\bf 9}(4), 045002
  (2023).

\bibitem{Cou23}
{Courtney-Barrer, B.}, {De Rosa, R.}, {Kokotanekova, R.}, {\em et~al.},
  ``Empirical contrast model for high-contrast imaging a vlt/sphere case
  study,'' {\em A\&A} {\bf 680}, A34  (2023).

\bibitem{Gir22}
J.~H. Girard, J.~Leisenring, J.~Kammerer, {\em et~al.}, ``{JWST/NIRCam
  coronagraphy: commissioning and first on-sky results},'' in {\em Space
  Telescopes and Instrumentation 2022: Optical, Infrared, and Millimeter Wave},
   L.~E. Coyle, S.~Matsuura, and M.~D. Perrin, Eds.,  {\bf 12180}, 121803Q,
  International Society for Optics and Photonics, SPIE  (2022).

\bibitem{Men23}
C.~B. Mendillo, K.~Hewawasam, J.~Martel, {\em et~al.}, ``{Balloon flight
  demonstration of coronagraph focal plane wavefront correction with
  PICTURE-C},'' {\em Journal of Astronomical Telescopes, Instruments, and
  Systems} {\bf 9}(2), 025005  (2023).

\bibitem{Mor22}
R.~E. Morgan, S.~Vlahakis, E.~Douglas, {\em et~al.}, ``{On-orbit operations
  summary for the Deformable Mirror Demonstration Mission (DeMi) CubeSat},'' in
  {\em Adaptive Optics Systems VIII},  L.~Schreiber, D.~Schmidt, and E.~Vernet,
  Eds.,  {\bf 12185}, 121857O, International Society for Optics and Photonics,
  SPIE  (2022).

\bibitem{Cad16}
E.~{Cady}, C.~M. {Prada}, X.~{An}, {\em et~al.}, ``{Demonstration of high
  contrast with an obscured aperture with the WFIRST-AFTA shaped pupil
  coronagraph},'' {\em Journal of Astronomical Telescopes, Instruments, and
  Systems} {\bf 2}, 011004  (2016).

\bibitem{Seo16}
B.-J. {Seo}, B.~{Gordon}, B.~{Kern}, {\em et~al.}, ``{Hybrid Lyot coronagraph
  for wide-field infrared survey telescope-astrophysics focused telescope
  assets: occulter fabrication and high contrast narrowband testbed
  demonstration},'' {\em Journal of Astronomical Telescopes, Instruments, and
  Systems} {\bf 2}, 011019  (2016).

\bibitem{Cad17}
E.~{Cady}, K.~{Balasubramanian}, J.~{Gersh-Range}, {\em et~al.}, ``{Shaped
  pupil coronagraphy for WFIRST: high-contrast broadband testbed
  demonstration},'' in {\em Techniques and Instrumentation for Detection of
  Exoplanets VIII},  {\em Proceedings of the SPIE} {\bf 10400}, 10400--14
  (2017).

\bibitem{Seo17}
B.-J. {Seo}, E.~{Cady}, B.~{Gordon}, {\em et~al.}, ``{Hybrid Lyot coronagraph
  for WFIRST: high-contrast broadband testbed demonstration},'' in {\em
  Techniques and Instrumentation for Detection of Exoplanets VIII},  {\em
  Proceedings of the SPIE} {\bf 10400}, 10400--15  (2017).

\bibitem{Mar17}
D.~{Marx}, B.-J. {Seo}, E.~{Sidick}, {\em et~al.}, ``{Electric field
  conjugation in the presence of model uncertainty},'' in {\em Techniques and
  Instrumentation for Detection of Exoplanets VIII},  {\em Proceedings of the
  SPIE} {\bf 10400}, 10400--23  (2017).

\bibitem{Rig18}
A.~J.~E. Riggs, G.~Ruane, E.~Sidick, {\em et~al.}, ``{Fast linearized
  coronagraph optimizer (FALCO) I: a software toolbox for rapid coronagraphic
  design and wavefront correction},'' in {\em Space Telescopes and
  Instrumentation 2018: Optical, Infrared, and Millimeter Wave},  M.~Lystrup,
  H.~A. MacEwen, G.~G. Fazio, {\em et~al.}, Eds.,  {\bf 10698}, 106982V,
  International Society for Optics and Photonics, SPIE  (2018).

\bibitem{Giv11}
A.~{Give'on}, B.~D. {Kern}, and S.~{Shaklan}, ``{Pair-wise, deformable mirror,
  image plane-based diversity electric field estimation for high contrast
  coronagraphy},'' in {\em Society of Photo-Optical Instrumentation Engineers
  (SPIE) Conference Series},  {\em Society of Photo-Optical Instrumentation
  Engineers (SPIE) Conference Series} {\bf 8151}  (2011).

\bibitem{Giv07}
A.~{Give'on}, B.~{Kern}, S.~{Shaklan}, {\em et~al.}, ``{Broadband wavefront
  correction algorithm for high-contrast imaging systems},'' in {\em Society of
  Photo-Optical Instrumentation Engineers (SPIE) Conference Series},  {\em
  Society of Photo-Optical Instrumentation Engineers (SPIE) Conference Series}
  {\bf 6691}  (2007).

\bibitem{Gro15}
T.~D. Groff, A.~J.~E. Riggs, B.~Kern, {\em et~al.}, ``{Methods and limitations
  of focal plane sensing, estimation, and control in high-contrast imaging},''
  {\em Journal of Astronomical Telescopes, Instruments, and Systems} {\bf
  2}(1), 011009  (2015).

\bibitem{Ahn23}
K.~{Ahn}, O.~{Guyon}, J.~{Lozi}, {\em et~al.}, ``{Combining EFC with spatial
  LDFC for high-contrast imaging on Subaru/SCExAO},'' {\em A\&A} {\bf 673}, A29
   (2023).

\bibitem{Pot20}
A.~{Potier}, R.~{Galicher}, P.~{Baudoz}, {\em et~al.}, ``{Increasing the raw
  contrast of VLT/SPHERE with the dark hole technique. I. Simulations and
  validation on the internal source},'' {\em A\&A} {\bf 638}, A117  (2020).

\bibitem{Van24}
K.~V. Gorkom, E.~S. Douglas, K.~Milani, {\em et~al.}, ``{The space coronagraph
  optical bench (SCoOB): 4. Vacuum performance of a high contrast imaging
  testbed},'' in {\em Space Telescopes and Instrumentation 2024: Optical,
  Infrared, and Millimeter Wave},  L.~E. Coyle, S.~Matsuura, and M.~D. Perrin,
  Eds.,  {\bf 13092}, 1309222, International Society for Optics and Photonics,
  SPIE  (2024).

\bibitem{Red21}
S.~F. Redmond, L.~Pueyo, L.~Pogorelyuk, {\em et~al.}, ``{Implementation of a
  broadband focal plane estimator for high-contrast dark zones},'' in {\em
  Techniques and Instrumentation for Detection of Exoplanets X},  S.~B. Shaklan
  and G.~J. Ruane, Eds.,  {\bf 11823}, 118231Q, International Society for
  Optics and Photonics, SPIE  (2021).

\bibitem{Zho16}
H.~Zhou, J.~Krist, and B.~Nemati, ``{Diffraction modeling of finite subband EFC
  probing on dark hole contrast with WFIRST-CGI shaped pupil coronagraph},'' in
  {\em Modeling, Systems Engineering, and Project Management for Astronomy
  VII},  G.~Z. Angeli and P.~Dierickx, Eds.,  {\bf 9911}, 99111S, International
  Society for Optics and Photonics, SPIE  (2016).

\bibitem{Goo96}
J.~Goodman, {\em Introduction to Fourier Optics}, McGraw-Hill  (1996).

\bibitem{Sou07}
R.~Soummer, L.~Pueyo, A.~Sivaramakrishnan, {\em et~al.}, ``Fast computation of
  lyot-style coronagraph propagation,'' {\em Opt. Express} {\bf 15}(24),
  15935--15951  (2007).

\bibitem{Rig23}
A.~J.~E. Riggs, M.~Bertagna, G.~J. Ruane, {\em et~al.}, ``{Coralign: a software
  package for coronagraphic alignment and calibration},'' in {\em Techniques
  and Instrumentation for Detection of Exoplanets XI},  G.~J. Ruane, Ed.,  {\bf
  12680}, 126802F, International Society for Optics and Photonics, SPIE
  (2023).

\bibitem{Zho18}
H.~Zhou, J.~Krist, E.~Cady, {\em et~al.}, ``{High accuracy coronagraph flight
  WFC model for WFIRST-CGI raw contrast sensitivity analysis},'' in {\em Space
  Telescopes and Instrumentation 2018: Optical, Infrared, and Millimeter Wave},
   M.~Lystrup, H.~A. MacEwen, G.~G. Fazio, {\em et~al.}, Eds.,  {\bf 10698},
  106982M, International Society for Optics and Photonics, SPIE  (2018).

\bibitem{Llo22}
J.~Llop-Sayson, A.~J.~E. Riggs, D.~P. Mawet, {\em et~al.}, ``{Coronagraph
  design with the electric field conjugation algorithm},'' {\em Journal of
  Astronomical Telescopes, Instruments, and Systems} {\bf 8}(1), 015003
  (2022).

\bibitem{Wil21}
S.~D. Will, T.~D. Groff, and J.~R. Fienup, ``{Jacobian-free coronagraphic
  wavefront control using nonlinear optimization},'' {\em Journal of
  Astronomical Telescopes, Instruments, and Systems} {\bf 7}(1), 019002
  (2021).

\bibitem{Sid17}
E.~Sidick, B.-J. Seo, B.~Kern, {\em et~al.}, ``{Optimizing the regularization
  in broadband wavefront control algorithm for WFIRST coronagraph},'' in {\em
  Techniques and Instrumentation for Detection of Exoplanets VIII},
  S.~Shaklan, Ed.,  {\bf 10400}, 1040022, International Society for Optics and
  Photonics, SPIE  (2017).

\bibitem{Rua22}
G.~Ruane, A.~J.~E. Riggs, E.~Serabyn, {\em et~al.}, ``{Broadband vector vortex
  coronagraph testing at NASA's high contrast imaging testbed facility},'' in
  {\em Space Telescopes and Instrumentation 2022: Optical, Infrared, and
  Millimeter Wave},  L.~E. Coyle, S.~Matsuura, and M.~D. Perrin, Eds.,  {\bf
  12180}, 1218024, International Society for Optics and Photonics, SPIE
  (2022).

\bibitem{Gol13}
G.~{Golub} and C.~{Van Loan}, {\em Matrix Computations}, The Johns Hopkins
  University Press  (2013).

\bibitem{Shi17a}
F.~Shi, E.~Cady, B.-J. Seo, {\em et~al.}, ``{Dynamic testbed demonstration of
  WFIRST coronagraph low order wavefront sensing and control (LOWFS/C)},'' in
  {\em Techniques and Instrumentation for Detection of Exoplanets VIII},
  S.~Shaklan, Ed.,  {\bf 10400}, 104000D, International Society for Optics and
  Photonics, SPIE  (2017).

\bibitem{Zho19}
H.~Zhou, J.~E. Krist, B.~D. Kern, {\em et~al.}, ``{WFIRST Phase B HLC occulter
  mask baselining and testbed WFC performance validation},'' in {\em Techniques
  and Instrumentation for Detection of Exoplanets IX},  S.~B. Shaklan, Ed.,
  {\bf 11117}, 111170H, International Society for Optics and Photonics, SPIE
  (2019).

\bibitem{Pog22}
L.~Pogorelyuk, C.~Haughwout, N.~Belsten, {\em et~al.}, ``{Computational
  complexities of image plane algorithms for high contrast imaging in space
  telescopes},'' {\em Journal of Astronomical Telescopes, Instruments, and
  Systems} {\bf 8}(4), 049003  (2022).

\bibitem{Bel23}
N.~Belsten, K.~Milani, L.~Pogorelyuk, {\em et~al.}, ``{Evaluating embedded
  hardware for high-order wavefront sensing and control},'' in {\em Techniques
  and Instrumentation for Detection of Exoplanets XI},  G.~J. Ruane, Ed.,  {\bf
  12680}, 126801N, International Society for Optics and Photonics, SPIE
  (2023).

\bibitem{Bai23}
V.~P. Bailey, E.~Bendek, B.~Monacelli, {\em et~al.}, ``{Nancy Grace Roman Space
  Telescope coronagraph instrument overview and status},'' in {\em Techniques
  and Instrumentation for Detection of Exoplanets XI},  G.~J. Ruane, Ed.,  {\bf
  12680}, 126800T, International Society for Optics and Photonics, SPIE
  (2023).

\bibitem{Tra16}
J.~T. Trauger, D.~C. Moody, J.~E. Krist, {\em et~al.}, ``{Hybrid Lyot
  coronagraph for WFIRST-AFTA: coronagraph design and performance metrics},''
  {\em Journal of Astronomical Telescopes, Instruments, and Systems} {\bf
  2}(1), 011013  (2016).

\bibitem{Zim16a}
N.~T. {Zimmerman}, A.~J.~E. {Riggs}, N.~J. {Kasdin}, {\em et~al.}, ``{Shaped
  pupil Lyot coronagraphs: high-contrast solutions for restricted focal
  planes},'' {\em Journal of Astronomical Telescopes, Instruments, and Systems}
  {\bf 2}, 011012  (2016).

\bibitem{Rig21}
A.~J.~E. Riggs, V.~Bailey, D.~C. Moody, {\em et~al.}, ``{Flight mask designs of
  the Roman Space Telescope coronagraph instrument},'' in {\em Techniques and
  Instrumentation for Detection of Exoplanets X},  S.~B. Shaklan and G.~J.
  Ruane, Eds.,  {\bf 11823}, 118231Y, International Society for Optics and
  Photonics, SPIE  (2021).

\bibitem{Ygo21}
M.~Ygouf, N.~Zimmerman, V.~Bailey, {\em et~al.}, ``Roman coronagraph instrument
  post processing report - os9 hlc distribution.''
  https://roman.ipac.caltech.edu/docs/sims/
  20210402\_Roman\_CGI\_post\_processing\_report\_URS.pdf  (2021).

\bibitem{Zel22}
R.~T. Zellem, B.~Nemati, G.~Gonzalez, {\em et~al.}, ``{Nancy Grace Roman Space
  Telescope coronagraph instrument observation calibration plan},'' in {\em
  Space Telescopes and Instrumentation 2022: Optical, Infrared, and Millimeter
  Wave},  L.~E. Coyle, S.~Matsuura, and M.~D. Perrin, Eds.,  {\bf 12180},
  121801Z, International Society for Optics and Photonics, SPIE  (2022).

\bibitem{Cad14}
E.~{Cady} and S.~{Shaklan}, ``{Measurements of incoherent light and background
  structure at exo-Earth detection levels in the High Contrast Imaging
  Testbed},'' in {\em Space Telescopes and Instrumentation 2014: Optical,
  Infrared, and Millimeter Wave},  {\em Proceedings of the SPIE} {\bf 9143},
  914338  (2014).

\end{thebibliography}
\bibliographystyle{spiejour}

\vspace{2ex}\noindent\textbf{Eric Cady} is an optical engineer in the high-contrast imaging group at NASA's Jet Propulsion Laboratory, and the lead for High-order Wavefront Sensing and Control (HOWFSC) on the Coronagraph Instrument (CGI) for the Nancy Grace Roman Space Telescope.  His research interests include coronagraphy, wavefront control systems, and starshades.  He received his B.S. in physics from Caltech in 2005 and a Ph.D. in mechanical and aerospace engineering from Princeton University in 2010.

\vspace{2ex}\noindent\textbf{Ilya Poberezhskiy} is a systems engineer at NASA's Jet Propulsion Laboratory, working on development and testing of optical instruments and subsystems for space applications. On the Roman Coronagraph Instrument project, he initially led the technology development team that demonstrated coronagraph technology readiness for flight implementation, and later served as the instrument project systems engineer. He received his B.S, M.S., and Ph.D. in electrical engineering from the University of California, Los Angeles (UCLA).

\vspace{2ex}\noindent\textbf{A J Eldorado (A.J.) Riggs} is an optical engineer at NASA's Jet Propulsion Laboratory. His research focuses on the high-contrast imaging of exoplanets, in particular mask optimization and wavefront sensing and control for the Coronagraph Instrument on the Nancy Grace Roman Space Telescope. He received his B.S. in physics and mechanical engineering from Yale University in 2011 and his Ph.D. in mechanical and aerospace engineering from Princeton University in 2016.

\vspace{2ex}\noindent\textbf{Byoung-Joon Seo} is an optical engineer at NASA's Jet Propulsion Laboratory. He has been working on the Coronagraph instrument (CGI) for Nancy Roman Space telescope (RST), coronagraph testbeds for Exoplanet Exploration Program (ExEP), alignment algorithm development for the James Webb Space Telescope (JWST), and optical modeling for Thirty Meter Telescope (TMT). He received a BS from Seoul National University, South Korea and MS/Ph.D from UCLA Electrical Engineering. 

\vspace{1ex}
\noindent Biographies of the other authors are not available.

\listoffigures
\listoftables

\end{document}